\documentclass[11pt,a4paper]{article}
\pdfoutput=1
\usepackage{jheppub}
\usepackage[utf8]{inputenc}
\usepackage{enumitem}
\usepackage{slashed}

\author{Christian Copetti}

\affiliation{Instituto de F\'isica Te\'orica UAM/CSIC, c/Nicol\'as Cabrera 13-15, Universidad Aut\'onoma de Madrid, Cantoblanco, 28049 Madrid, Spain}

\emailAdd{christian.copetti@uam.es}

\title{Torsion and anomalies in the warped limit of Lifschitz theories}	
\date{\today}

\abstract{
We describe the physics of fermionic Lifschitz theories once the anisotropic scaling exponent is made arbitrarily small. In this limit the system acquires an enhanced (Carrollian) boost symmetry. We show, both through the explicit computation of the path integral Jacobian and through the solution of the Wess-Zumino consistency conditions, that the translation symmetry in the anisotropic direction becomes anomalous. This turns out to be a mixed anomaly between boosts and translations. In a Newton-Cartan formulation of the space-time geometry such anomaly is sourced by torsion. We use these results to give an effective field theory description of the anomalous transport coefficients, which were originally computed through Kubo formulas in \cite{copetti2019anomalous}. Along the way we provide a link with warped CFTs.}

\begin{document}

\maketitle

\flushbottom

\pagebreak

\section*{Introduction}
\addcontentsline{toc}{section}{Introduction}
Quantum critical points and their physics have attracted a huge amount of interest over the last decade \cite{sachdev2007quantum}. In this paper we will focus on a nonrelativistic class of such theories and their effective description at low energies. We will focus on fermionic theories with an emergent Lifschitz scaling symmetry. A well known example is given by the low energy limit of the following four dimensional Lagrangian
\begin{equation}
\mathcal{L}= \bar{\psi} \left( i \gamma^\mu \partial_\mu -m + \gamma^\mu \gamma_5 n_\mu \right) \psi \, .
\end{equation}
This is often interpreted, in condensed matter language \cite{RevModPhys.90.015001}, as describing the transition between a Weyl semimetal and a trivial insulator. The quantum critical point is reached upon tuning $|m|=|n|$ (we take $n_\mu$ to be a spatial vector) and its low energy excitations are characterized by an emergent anisotropic Lifschitz scaling symmetry with $z=1/2$. This can be seen from the dispersion relation at criticality
\begin{equation}
\epsilon^2(k) = k_{\perp}^2 + \frac{1}{4 m^2} k_v^4 + ... \, ,
\end{equation} 
with $k_v = k_\mu v^\mu$ and $n_\mu v^\mu=1$. The Lifschitz symmetry is in this case a bit unconventional, since it scales anisotropically a space-like direction instead of a time-like one by $x_a \to \lambda x_a$ and $x_v \to \lambda^z x_v$.\footnote{We find it conventient to introduce vector fields $v^\mu$, $E^\mu_a$ that satisfy $n_\mu v^\mu=0$ , $E^\mu_a e_\mu^b =\delta_{a}^b$, with all other contractions vanishing, to decompose $x^\mu= x_v v^\mu + x^a E_a^\mu$. Such a decomposition is always subtended by this notation and is familiar in Newton Cartan geometry. In Section \ref{Sec:Carroll} this is clarified in the framework of Carrollian geometry.} In the case at hand $z=1/2$, however the values $z=1/2N$, for integer $N$, may be reached by adding $N$ fine-tuned couplings to higher spin chiral currents.\footnote{See Appendix A of \cite{copetti2019anomalous}.}\newline
To describe the physics of the critical point one may employ the following action for the fermionic fluctuations
\begin{equation}
S= \int d^{d}x \sqrt{g} \left( i \bar{\varphi} \gamma^a E^\mu_a \nabla_\mu \varphi + s \bar{\varphi} M_z \varphi\right) \, , \label{Intro:effaction0}
\end{equation}  
with $d=2,4$, $ M_z = \left(i \nabla_v \right)^{1/z}$ and $E^\mu_a$ the (reduced) inverse vielbein. From here on $\gamma^a$ are Clifford matrices of the $SO(1,d-1)$ algebra (or $SO(d-1)$ in Eulidean signature) while  $s=\pm$ is a book-keeping parameter odd under time reversal. In some cases it is convenient to think of the model as a towe of lower dimensional Dirac fermion with $k_v$ dependent mass $M_z$. The Lifschitz symmetry acts as
\begin{equation}
x_a \to \lambda \ x_a \, , \ \ x_v \to \lambda^z \ x_v \, \ \ \varphi \to \lambda^{-(d-2+z)/2} \ \varphi \, .
\end{equation}
An important observation is that this model has a marginal coupling, given by the normalization of $M_z$, the anisotropic part of the kinetic term. To be more precise, the Lifschitz invariant quantity is the ratio between the isotropic and anisotropic parts of the kinetic terms, we choose to normalize the isotropic kinetic term to one for ease of notation. It is convenient to introduce this coupling by replacing $s \to s q$ and $\nabla_v \to \nabla_v/q$, in this way $M_z= s q \left(\frac{i \nabla_v}{q^2} \right)^{1/z}$ and, if one wishes to adopt the standard dimensional counting in which all derivatives have the same scaling, $q$ can be interpreted as a physical momentum scale. We however stress that, from the Lifschitz perspective, $q$ is dimensionless. A useful way to rephrase these observations is to say thath the system is invariant under the spurionic symmetry generated by 
\begin{equation}
x_\mu \to r \ x_\mu \ , \ \ q \to r^{-1} \ q \, , \ \ \varphi \to r^{-(d-1)/2} \ \varphi . \label{eq:spurionic}
\end{equation}
This can be used to fix the power of $q$ that appears in most physical quantities.\newline

\noindent The main difference between two and four dimensions are the allowed valued of $z$. We impose that the critical theory breaks time reversal and is local. This leads in two dimensions to $z=1/(2N+1)$, $N \in \mathbb{N}$, while in four dimensions to $z=1/2N$. With these conventions $M_z$ contains no fractional power of derivatives and the theory is local. Notice however that this will still allow us to examine the limit $z \to 0$ without referring to non-local theories.\\

\begin{figure}
	\centering
	\includegraphics[width=1.0\linewidth]{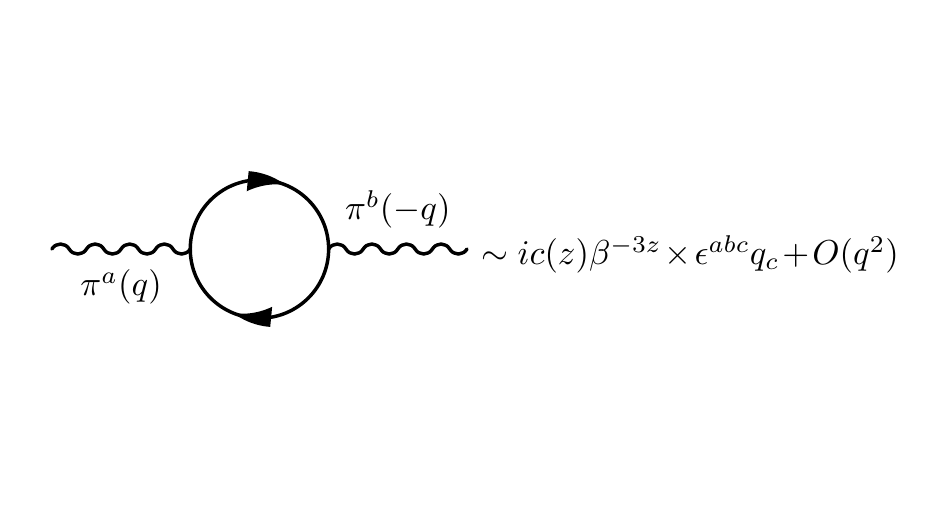}
	\caption{One-loop computation of the Hall contribution to the anisotropic momentum current. This leads to the torsional response since, in the Newton-Cartan geometry, $T_{\mu\nu}=--\partial_{[\mu} n_{\nu]}$}
	\label{Fig1}
\end{figure}

\noindent Since the model is invariant under charge conjugation, one can furthermore impose the Majorana condition $\varphi^* = \varphi$ on the fermion, which gives the precise model studied in \cite{copetti2019anomalous}
\begin{equation}
S= \int d^{d}x \sqrt{g} \left( i \bar{\varphi} \gamma^a E^\mu_a \nabla_\mu \varphi + s \varphi^T \mathcal{C}^{-1} M_z \varphi\right) \, , \label{Intro:effaction} \, ,
\end{equation}
with $\mathcal{C}$ the charge conjugation matrix. \newline
These models have been show both holographically \cite{Landsteiner:2016stv} and by field theory methods \cite{copetti2019anomalous} to possess a distinctive nontrivial response at finite temperature (for discussions of other regimes see e.g. \cite{Link:2017ora,Pena-Benitez:2018dar,Offertaler:2018gwz}). More precisely, these models conserve momentum in the anisotropic direction $x_v$. This gives rise by Noether's theorem to a conserved current $\pi^\mu\equiv \frac{1}{\sqrt{g}} \frac{\delta S}{\delta n_\mu}$. The main result of \cite{copetti2019anomalous} is that the two point function of two such currents, represented in Figure \ref{Fig1} is nonzero in the Kubo limit. Using basic tools of Newton Cartan geometry \cite{Duval:2009vt,Jensen:2014aia}, it can be shown that this corresponds to a nontrivial response to external torsion $T_{\mu\nu}=- \partial_{[\mu}n_{\nu]}$ of the form
\begin{equation}
\pi^\mu = s q^3 \beta^{-3 z} c(z) \epsilon^{\mu\nu\rho\sigma} n_\nu T_{\rho\sigma}\, ,
\end{equation}
with $\beta$ the inverse temperature and $c(z)$ a non-universal function of $z$. What will be of importance for us is that $c(z)$ has a nontrivial limit for $z \to 0$ for which the result is $\beta$-independent:
\begin{equation}
\lim_{z \to 0} \pi^\mu = \frac{s q^3}{48 \pi^2} \epsilon^{\mu\nu\rho\sigma} n_\nu T_{\rho\sigma} \, . 
\end{equation}
In \cite{copetti2019anomalous} these effects where given an effective field theory interpretation using Chern-Simons theory, by integrating out the massive modes for which $M_z\gg 1$ and dimensionally reducing along the anisotropic direction. In this respenct the theory is similar to the free massive Dirac fermion in 2+1 dimensions, where the fermion can be integrated out generating an effective Chern-Simons description of the $U(1)$ dynamics.\\

\noindent Since in many cases (e.g. the chiral magnetic effect \cite{Son:2009tf,Kharzeev:2010gd,Kharzeev:2013ffa}) such non-dissipative effects can be explained in a universal way by studying the 't Hooft anomalies of the theory, in this paper we develop such perspective for the Lifschitz model \eqref{Intro:effaction}. In particular, we construct the effective description of the warped limit $z \to 0$ generalizing the results of \cite{jensen2017locality} to four dimensions, pointing out a way to reinterpret the interesting low energy observables in the framework of chiral physics. This extends various results from the existing literature (see e.g. \cite{Landsteiner:2016led} for a review) to warped systems.\\

\noindent There are two ways to analyze the warped limit. One is to compute in the Lifschitz theory and then take the limit of $z \to 0$. In this case the simplest characterization of anomalies comes from evaluating the change in the fermionic Jacobian by the Fujikawa procedure \cite{fujikawa1979path} and extract the finite terms as the warped limit is approached. 
The other approach is to directly work in the warped limit. The basic geometric framework to do so was developed by various authors \cite{hofman2011chiral,hofman2015warped,jensen2017locality} in the context of warped conformal field theories. These are scale invariant theories whose algebra contains additional ``Carrollian'' boost generators. To be precise the non-vanishing commutators are:
\begin{align}
[J_{ab}, J_{cd}] &= \eta_{ac} J_{bd} - \eta_{ad} J_{bc} + \eta_{bd} J_{ac} - \eta_{bc} J_{ad} \, , \\
[J_{ab},P_c]&= \eta_{ac} P_b - \eta_{bc}P_a \, \ \ [J_{ab},C_c]= \eta_{ac} C_b - \eta_{bc}C_a \, , \\
[P_a, P_b] &=0 \, , \ \ \ \ [C_a,C_b]=0 \, \\
[P_a,C_b]&= \eta_{ab} \Pi \, , 
\end{align}
Where $P_a$ are the (isotropic) momenta, $C_a$ the boosts, $\Pi$ the anisotropic momentum, $J_{ab}$ the rotation generators. These are complemented by the warped Lifschitz scaling
\begin{align}
[D,J_{ab}]=0 \, , \ \ [D,P_a]&= -P_a \, \ \ [D,C_a]=C_a \, , \\
[D,\Pi]&= 0 \, . 
\end{align}
\noindent The presence of boost is important and non-trivial. We can show that indeed the fermionic theory \eqref{Intro:effaction0} satisfies the right boost Ward identity at the classical level in the warped limit, so that its symmetry enhances. By coupling our system to a curved background geometry, one can consistently ask about its anomalies by solving the Wess-Zumino consistency conditions \cite{Wess:1971yu}. The Fujikawa analysis in this case presents difficulties in defining appropriate regulators, probably due to the somewhat singular nature of the warped limit.\\

\noindent In the body of the paper we perform the analyisis in both ways indicated above and find perfect agreement in their predictions. It has to be noted that the descent equations do not fix the anomaly coefficients, in this sense the Lifschitz computations serves as a microscopic input, while the descent procedure essentially proves regulator-independence. The final conclusion is that, once torsion is included, it leads to an anomaly in the Ward identities for the current $\pi^\mu$ of the form
\begin{equation}
\frac{1}{\sqrt{g}} \partial_\mu \sqrt{g} \pi^\mu = \kappa \epsilon^{\mu\nu\rho\sigma} T_{\mu\nu} T_{\rho\sigma} \, ,
\end{equation}
with, in particular
\begin{equation}
\kappa = \frac{q^3}{32 \pi^2} \, ,
\end{equation}
for the fermionic models considered.\\

\noindent The paper is organized as follows
\begin{description}
	\item [Section 1] We start by describing the physics of the Lifchitz theories \eqref{Intro:effaction}, showing that, upon Fujikawa regularization, they have a nontrivial anomaly in the anisotropic translations as the warped limit is taken. This anomaly is sourced by torsion, it does not come, however, with a divergent UV cutoff, as it is customary when torsion appears. This state of affairs is ultimately a consequence of the warped conformal invariance. It can be used to derive an analogue of the chiral magnetic effect for warped Lifschitz theories, of which we give details in Appendix \ref{App:Kubo}. 
	\item [Section 2] We introduce the natural (Carrollian) geometries \cite{hartong2015gauging,detournay2012warped,duval2014carroll} in which to study warped theories. We use this setting to give solutions to the Wess Zumino consistency conditions \cite{Bardeen:1984pm}. Once a set of curvature constraints is imposed, we show that there is an emergent Stueckelberg field which allows for nontrivial solutions to exist. These match the Lifschitz predictions. 
		\item  [Section 3] We introduce warped conformal field theories (WCFTs) and their free field realizations. We prove that the Lifschitz system \eqref{Intro:effaction0} reduces to one such theory in the warped limit and match their marginal coupling and relevant Ward identities. Finally we analyze free warped theories in various dimensions, pointing out a mechanism to convert the ``space-time" translation symmetry in the anisotropic direction in an internal chiral symmetry. This allows for a simple interpretation of the previous results.
\end{description}
	Finally, we conclude with open questions and remarks. Technical or lengthy passages instrumental to proving the main results are summarized in various Appendices.
\section{The Lifschitz fermion}\label{Sec: LIfschitz}
In this section we study the Fujikawa regularization of translations in the anisotropic direction for the action \eqref{Intro:effaction}, which for clarity we racall here
\begin{equation}
S= \int d^{d}x \sqrt{g} \left( i \bar{\varphi} \gamma^a E^\mu_a \nabla_\mu \varphi + s q \varphi^T \mathcal{C}^{-1} M_z \varphi\right) \, ,
\end{equation}
with $M_z = \left(\nabla_v/q\right)^{1/z}$. To find the anomalous variation of the effective action we first need to couple this system to the relevant external gauge fields. In our case this means introducing the appropriate background geometry for the Lifschitz system. In \cite{copetti2019anomalous} this was realized by considering a Newton-Cartan setup without any boost symmetry nor $U(1)$ gauge field. We briefly review it here. It amounts to the geometrical data $n_\mu$, $e_\mu^a$ with their algebraic inverses $v^\mu$, $E^\mu_a$ satisfying
\begin{equation}
n_\mu v^\mu =1 \, , \ \ E^\mu_a e_\mu^b = \delta^b_a \, , \ \ e_\mu^a v^\mu = E^\mu_a n_\mu = 0 \, ,
\end{equation}
the connection is fixed by demanding 
\begin{equation}
\nabla_\mu n_\nu = \nabla_\mu v^\nu = \nabla_\mu e_\nu^a = \nabla_\mu E^\nu_a = 0 \, ,
\end{equation}
with $\nabla$ containing both the Christoffel symbols and the spin connection $\omega_\mu^{ab}$. One obtains the connection
\begin{equation}
\Gamma_{\mu\nu}^\rho = - v^\rho \partial_\mu n_\nu + \frac{1}{2}h^{\rho\lambda} \left(-\partial_\lambda h_{\mu\nu} + \partial_\nu h_{\mu\lambda} +\partial_\mu h_{\lambda\nu}\right) \, , \ \ h_{\mu\nu}= e_\mu^a e_\nu^b \eta_{ab} \, ,
\end{equation}
with torsion
\begin{equation}
T_{\mu\nu}= n_\rho \Gamma^\rho_{[\mu\nu]}= - \partial_{[\mu} n_{\nu]} \, , 
\end{equation}
and the further zero extrinsic curvature constraint $\mathcal{L}_v h_{\mu\nu}=0$. The spin connection has the usual form in terms of the vielbein $e_\mu^a$. In this geometry one may decompose any vector field $\xi^\mu = \theta v^\mu + \xi^a E^\mu_a$, with $\theta \, , \xi^a$ well defined parmaters due to the absence of boost symmetry. We are interested to compute the (regulated) effect of a nonzero $\theta$ in a diffeomorphism transformation for the Lifschitz system, we will call this transformation an ``anisotropic translation"\footnote{When dealing with systems with a boost symmetry we will use a first order formulation in which the generator of translations in the anisotropic direction enters as a gauge field. In that case the anisotropic translations are the analogue of P-translations of \cite{bergshoeff2019lie}. }. Since there is only a reduced rotation symmetry relating the various components of the stress tensor, it is useful to introduce the currents $\pi^\mu$ and $t^\mu_a$ as
\begin{align}
\pi^\mu &= \frac{1}{\sqrt{-g}} \frac{\delta S}{\delta n_\mu} \, \\
t^\mu_a &= \frac{1}{\sqrt{-g}} \frac{\delta S}{\delta e_\mu^a} \, ,
\end{align}
whose conservation is related to invariance of the action under diffeomorphisms generated by $\theta$ and $\xi^a$ respectively. To be precise, in the geometry we are using, the classical Ward identities read
\begin{align}
\left(\nabla_\mu -G_\mu\right) {t^\mu}_a - E_a^\nu T_{\nu\mu} \pi^\mu &= 0\, , \\
\left(\nabla_\mu -2 G_\mu\right)\pi^\mu &= 0 \, ,
\end{align}
with $G_\mu= v^\nu T_{\nu\mu}$. We will be interested in understanding how the Ward identity for $\pi^\mu$ may be violated in the presence of nontrivial background torsion.
\subsection{Fujikawa regularization for anisotropic translations}\label{sec:Fujikawa}
Torsional contributions to the anomaly polynomial of chiral field theories are not new, see for example \cite{chandia1997topological}. However, since the vielbein does not have the right scaling dimension for a connection, one concludes that torsion must always enter together with a UV scale $\Lambda$. For chiral theories, computing the change in the fermionic measure due to a chiral transformation in the presence of torsion, gives a term proportional to the Nie-Yan density
\begin{equation}
\Lambda^2 NY[e]= \Lambda^2 \int \left( T^a \wedge T_a - R_{ab} \wedge e^a \wedge e^b \right) \, , \label{Fuji:NY}
\end{equation} 
where $\Lambda$ is the UV cutoff introduced to regulate the divergent Jacobian. This is divergent and should be made to vanish by introducing appropriate counterterms. Such a problem is ubiquitous in theories with torsion, which has made the interpretation of statements like \eqref{Fuji:NY} controversial. The same problem will extend to Lifschitz systems, where however the field $n_\mu$ has dimensions $\Lambda^z$. One thus may hope that tuning $z$ to zero may give rise to finite contributions. A second point of view, advocated for example in \cite{hughes2011torsional,hughes2013torsional}, is that \eqref{Fuji:NY} should be interpreted as arising form the boundary modes of a gapped system with gap $\sim \Lambda$. The coefficient then can be regulated using Pauli-Villars fields. \newline
\noindent For fermionic theories the 't Hooft anomaly polynomial can be extracted via the well known Fujikawa procedure by explicitly computing the (regulated) change in the fermionic measure under a local symmetry transformation. To linear order this is just the trace of the field variation.\newline
To estimate these contributions we adopt a covariant regularization of the path integral Jacobian via heath kernel
\begin{equation}
\textbf{A}(\delta_\Psi)= \lim_{\Lambda \to \infty} tr \left[ \delta_\Psi e^{\mathcal{R}} \right] \, ,
\end{equation}   
with $\delta_\Psi$ the infinitesimal variation of the fermionic fields and $\mathcal{R}$ the covariant regulator. This will depend on a set of UV paramters $\Lambda$ which we take to be diverging as customary. The choice of $\mathcal{R}$ is dictated by the symmetries of the problem, although they do not fix it completely. One should impose:
\begin{itemize}
	\item $e^{\mathcal{R}}$ to have finite trace, that is to decay fast enough in all directions in momentum space.
	\item $\mathcal{R}$ to be covariant under Lifschitz transformations and invariant up to a rescaling of the $\Lambda$s.
	\item $\mathcal{R}$ needs to couple consistently to the background geometry.
	\item We will also assume the regulator respects the spurionic scaling symmetry \eqref{eq:spurionic}. This fixes how $q$ should appear. Notice that this means that the $\Lambda$s may not transform under such a symmetry.
\end{itemize}
The simplest candidate which satisfies these requirements is given by
\begin{equation}
\mathcal{R}= A^\dagger A \, ,
\end{equation}
with $A$ related to the Dirac operator as
\begin{equation}
A = \frac{i  \gamma^a \nabla_a/q}{\Lambda_1} + s  \frac{\left(i \nabla_v/q \right)^{1/z}}{\Lambda_2} \, ,
\end{equation}
where we introduce $s$ to keep track of the time-reversal transformation properties of the terms in the heath-kernel expansion. Notice that we have introduced two independent parameters $\Lambda_1$, $\Lambda_2$ since there is no rotation symmetry relating them. We will be interested in the limit in which both are taken to be large. More precisely we will take the $\Lambda$s to be large \emph{but} finite and take the limits $z \to 0$ and $\Lambda \to \infty$ in this order. The conditions listed above tell us that both parameters need to scale under Lifschitz transformations with weight one to preserve scale invariance, while they are neutral under the spurionic symmetry.

\noindent We then have to expand the regulator in external fields. In two dimensions $\gamma^a$ is just the identity matrix, thus the first term changes sign under Hermitean conjugation, in four dimensions $i \gamma^a$ is real in the Majorana representation, and it is the second term to change sign due to the odd number of $D_v$ derivatives. We then expand the regulator in curved spacetime by virtue of
\begin{equation}
[\nabla_\mu, \nabla_\nu]= - T_{\mu\nu} \nabla_v + R_{\mu\nu}^{ab} J_{ab} \, ,
\end{equation}
using standard manipulations this leads to
\begin{equation}
-\mathcal{R} = \frac{\nabla_{\perp}^2/q}{\Lambda_1^2} - \frac{i}{2 \Lambda_1^2}\epsilon^{abc}\gamma_a T_{bc} \nabla_v/q +i s \gamma^a \sum_k^{1/z} \frac{c_k}{\Lambda_1 \Lambda_2} (\nabla_v/q)^k G_a (\nabla_v/q)^{1/z-k} + \frac{(i \nabla_v/q)^{2/z}}{\Lambda_2^2} + \mathcal{R}(R) \, ,
\end{equation}
in four dimensions and
\begin{equation}
-\mathcal{R} = \frac{\nabla_{\perp}/q^2}{\Lambda_1^2}  +i s \sum_k^{1/z} \frac{c_k}{\Lambda_1 \Lambda_2} (\nabla_v/q)^k G_\mu E^\mu (\nabla_v/q)^{1/z-k} + \frac{(i \nabla_v/q)^{2/z}}{\Lambda_2^2} \, ,
\end{equation}
in two dimensions. We denote $\mathcal{R}(R)$ the contributions coming from the curvature of the spin connection. These may be relevant in describing mixed anomalies but we do not use them here. From now on they will be ignored. We have furthermore defined $G_\mu = T_{\mu\nu}v^\nu$ and $E^\mu$ as the 2d analogue of $E^\mu_a$. We also have the combinatorial coefficient $c_k= \sum_m^{1/z-k} {{m+k}\choose{k}}$. For anisotropic translations we have $\delta_\Psi = \theta \nabla_v$ and the regulated Jacobian
\begin{equation}
\textbf{A}(\theta)=  tr \left[ \theta \nabla_v e^{\mathcal{R}} \right]=  \int \frac{d^{d-1} k_a}{(2\pi)^{d-1}}\int \frac{d k_v}{(2\pi)} \theta (\nabla_v + i k_v)  e^{\mathcal{R}[\nabla +i k]} \, ,
\end{equation}
where we have chosen a plane wave basis for the expansion, see \cite{bastianelli2006path,fujikawa2004path,Bertlmann:1996xk} for more details. One then expands the integral around the Gaussian contribution order by order in $\Lambda_1$ and $\Lambda_2$, to be explicit
\begin{equation}
\textbf{A}(\theta)=\sum_{n_1, n_2} J^{n_1, n_2} (\theta) \, , J^{n_1, n_2}(\theta)\sim \Lambda_1^{n_1} \Lambda_2^{n_2} \, ,
\end{equation}
We will be interested in contributions that are $\mathcal{T}$ odd and are finite in the warped limit. The first condition just tells us that we must get something proportional to $s$, the second one that we want terms which go like $\Lambda_1^{f(z)} \Lambda_2^{g(z)}$, $\lim_{z\to 0} f(z)=\lim_{z\to 0} g(z)=0$. These two conditions together reduce strongly the number of terms that we need to consider.\newline
In two dimensions the expansion is very simple, since bringing down one factor of $G_\mu E^\mu$ does the job. Higher contributions are either $\mathcal{T}$ even or negligible. The relevant integral assumes a clearer physical form upon introducing dimensionless variables $k_a =q \ \Lambda_1  u_a$ and $k_v= q \ \Lambda_2^z v$. Then, using that $c_{1/z}=1/z$ 
\begin{equation}
\textbf{A}^{(2z)}(\theta) =\theta \frac{s q^2}{z (2\pi)^2} \Lambda_2^{2z} \int du dv \ v^{1+1/z} \exp(-u^2 - v^{2/z} )  E^\mu G_\mu \, ,  \label{Lif:intermediate}
\end{equation}
which leads upon integration to
\begin{equation}
\textbf{A}^{(2z)}(\theta) = \theta \frac{s q^2 }{4 \pi} \Lambda_2^{2z} \frac{\Gamma(z +1/2)}{\Gamma(1/2)} E^\mu G_\mu \, ,
\end{equation} 
the factor of $1/z$ in \eqref{Lif:intermediate} is adsorbed by the change of variables $v^2=x^z$. Using that in two dimensions $\epsilon^{\mu\nu} T_{\mu\nu}= 2 E^\mu G_\mu$ 
\begin{equation}
\lim_{z \to} \textbf{A}^{(2z)} (\theta)= J^{\rm warped}(\theta) = \theta \frac{q^2 s}{8\pi} \sqrt{g} \epsilon^{\mu\nu} T_{\mu\nu} \, ,
\end{equation}
notice that the final result is independent of $\Lambda_1$ and $\tilde{\Lambda}_2$, while the $q$ dependence is fixed by the spurionic symmetry. This should be confronted with the Jacobian for the chiral anomaly in 2d.\newline
In four dimensions one needs to expand to one order higher in the external torsion. To get a non-vanishing trace of $\gamma$ matrices we need once the term with $T_{ab}$ and once the one with $G_a$. In much the same way one gets
\begin{equation}
\textbf{A}^{(3z)}(\theta)= \theta \frac{s q^3 \Lambda_2^{3z}}{8 \pi^2} \frac{\Gamma(3z/2+1/2)}{\Gamma(1/2)} \epsilon^{abc}G_a T_{bc} \, ,
\end{equation}
and by virtue of 
\begin{equation}
\epsilon^{abc}G_a T_{bc}= \frac{1}{4} \epsilon^{\mu\nu\rho\sigma}T_{\mu\nu}T_{\rho\sigma}\, ,
\end{equation}
we find
\begin{equation}
\textbf{A}^{(3z)}_\theta =\theta \frac{s q^3 \Lambda_2^{3z}}{32 \pi^2} \frac{\Gamma(3z/2+1/2)}{\Gamma(1/2)} \epsilon^{\mu\nu\rho\sigma}T_{\mu\nu}T_{\rho\sigma} \, ,
\end{equation}
whose warped limit is
\begin{equation}
\textbf{A}^{\rm warped}(\theta) = \theta \frac{s q^3}{32 \pi^2} \epsilon^{\mu\nu\rho\sigma}T_{\mu\nu}T_{\rho\sigma} \, ,
\end{equation}
which should be confronted with the four dimensional chiral anomaly. It is interesting to notice that in this analogy the marginal coupling $q$ plays the role of the chiral charge. It should be appreciated that the breaking of time reversal in both dimensions is key to ensure nonvanishing integrals, that are otherwise cancelled between $k_v$ and $-k_v$.\newline 
In four dimensions it is known that a chiral system also has a mixed Lorentz/chiral anomaly in the presence of nontrivial curvature. The evaluation is however extremely cumbersome in our regularization scheme and, although such contribution does not seem to be present, one should resort to supersymmetric methods in order to give a definitive answer. The appearence of $q$ in the final result is entirely due to our choice of preserving the spurionic symmetry. We will see in fact that this indeed matches what is expected in the WCFT analysis. However, without imposing the spurionic symmetry from the get go, it is not clear how to match the marginal parameters in the two theories and the definition of $q$ in the warped limit may be ambiguous.\newline
We have thus shown the following 
\begin{itemize}
	\item The Lifschitz fermion displays a covariant anomaly sourced by torsion in the warped limit. The cutoff dependence of torsional terms vanishes as the one-form $n_\mu$ gets the right dimensions for a connection. We can use a wider class of regulators than in the isotropic case, however we have shown that the independent regulatory parameters vanish as $z \to 0$, leaving behind only powers of $q$ fixed by the spurionic symmetry. We show in Section \ref{Sec:Carroll} that no local counterterm can correct the result without making other symmetries anomalous.
	\item The resulting anomalies are extremely similar to the ones of a chiral system, where the chiral symmetry is somehow generated by anisotropic translations.
	\item The role of the chiral charge is taken over by the marginal coupling $q$. This number is not arbitrary, as we will show in the next Section that it fixes the anisotropic momentum of the only nontrivial modes in the warped limit. 
\end{itemize}
Admittedly our computation was somewhat cavalier, as we have disregarded subleading terms in the expansion of the Jacobian. To give further support for the arguments that we have given one may recall that chiral $U(1)$ systems display nontrivial response to chemical potentials in a magnetic field (e.g. the chiral magnetic effect). We have computed such contributions for the Lifschitz system in Appendix \ref{App:Kubo}. In our case the chemical potential is the one for anisotropic translations, which may also be interpreted as giving the system a nonvanishing velocity field in said direction. The results agree with those of \cite{copetti2019anomalous} once the right (covarianiant or consistent) currents are considered.

\section{Warped geometry and anomalies}\label{Sec:Carroll}
The Carroll group can be introduced as the $c \to 0$ contraction of the Poincaré algebra, much in the same way as the Galilei group is the $c \to \infty$ one. The geometry to which such a theory may be coupled and the dynamics of pointlike particles have been extensively studied in the past, see e.g. \cite{duval2014carroll,duval2014conformal,Banerjee:2014nja,bergshoeff2014dynamics,bergshoeff2017carroll,hartong2015gauging}. Here we present the main results that are needed to construct solutions of the Wess Zumino consistency conditions. We pay particular attention to the discussion of constraints that one may impose in gauging the Carroll algebra. We refer the reader to the references above for further deatils.\newline
The Carroll group is given by the following commutation relations
\begin{align}
[J_{ab}, J_{cd}] &= \eta_{ac} J_{bd} - \eta_{ad} J_{bc} + \eta_{bd} J_{ac} - \eta_{bc} J_{ad} \, , \\
[J_{ab},P_c]&= \eta_{ac} P_b - \eta_{bc}P_a \, \ \ [J_{ab},C_c]= \eta_{ac} C_b - \eta_{bc}C_a \, , \\
[P_a, P_b] &=0 \, , \ \ \ \ [C_a,C_b]=0 \, \\
[P_a,C_b]&= \eta_{ab} \Pi \, , 
\end{align}
where $J_{ab}$ generate the reduced rotation subgroup, $C_a$ the boosts, $P_a$ the ``isotropic" translations and $\Pi$ the anisotropic ones. We deviate slightly from the convention of denoting by $H$ the central term $\Pi$, since in our case it will not be the Hamiltonian of the system. This is an important point, since otherwise the dynamics of the system is completely trivial, being the energy of the particle completely fixed by the central element $H$. To this one might add a further scaling generator $D$ with
\begin{align}
[D,J_{ab}]=0 \, , \ \ [D,P_a]&= -P_a \, \ \ [D,C_a]=(1-z)C_a \, , \\
[D,\Pi]&= -z \Pi \, . 
\end{align}
We then see that the warped case $z=0$ is special, since it is the only one for which the algebra still has a central element\footnote{This is not true in $d=3$ where one can use the $\epsilon^{ab}$ tensor to introduce the following extensions
	\begin{equation}
	[P_a,P_b] \sim \epsilon^{ab} X \, , \ \ [C_a,C_b]\sim \epsilon^{ab} Y \, , \ \ [P_a,C_b]\sim \epsilon^{ab} Z \, ,
	\end{equation}
	while, with an eye to 2d edge modes, it could be interesting to discuss such situations, we will have nothing to say about them in this paper.
}. At the level of spacetime, one may check that Carrollian boosts act by sending the anisotropic coordinate $x_v$ to $x_v + \lambda_a x^a$ so that the algebra is consistent. \newline
The analysis of the anomaly polynomial can also in principle be done for other nonrelativistic theories (e.g. Galileian \cite{Jensen:2014hqa}). However in that case one expects anomalies to be only present in \emph{odd} dimensionality, due to the interpretation of Galileian theories as coming from null reductions. In the Carrollian case this is not a problem, since they can be seen as the theories which arise on null embeddings in Bargmann spacetime \cite{duval2014carroll,Bekaert:2015xua,Morand:2018tke,Donnay:2019jiz} and anomalies can be realized via the inflow mechanism. Furthermore, torsional anomalies in the Galileian case are not to be expected based on the dimensionality of the translations generators (minus one and minus two respectively). This should serve as a motivation to justify expecting a nontrivial result in the Carrollian case. \newline
 We now proceed to construct gauge theoretical description of Carrollian theories to set the ground for the determination of the possible anomaly polynomials. We will mainly follow \cite{hartong2015gauging}\footnote{However see also Appendix A of \cite{Bekaert:2015xua} and \cite{Bekaert:2014bwa}}.\newline
The first step in the gauging procedure is to introduce a Lie algebra valued connection $\mathcal{A}$, with the decomposition\footnote{It is important to stress that, despite of the nomenclature, on still cannot identify $n$ and $e^a$ with the vielbein and anisotropic one-form for the system, since for example, they do not span orthogonal directions. Such identifications come once one imposes that diffeomorphisms should be realized as gauge transformations of the connection $\mathcal{A}$.}
\begin{equation}
\mathcal{A}= n \Pi + e^a P_a + f^a C_a + \frac{1}{2}\omega^{ab} J_{ab} \, , \label{Geom:connection}
\end{equation}
to which one may associate a curvature 
\begin{equation}
\mathcal{F}= d \mathcal{A} + \frac{1}{2}[\mathcal{A},\mathcal{A}] =F(\Pi) \Pi + F(C)^aC_a + F(P)^a P_a  + \frac{1}{2}F(J)^{ab} J_{ab}\, ,
\end{equation}
whose components read explicitly
\begin{align}
F(\Pi)&= (dn - f^a\wedge e_a) \, , \\
F(C)^a &= D f^a \, , \\
F(P)^a &= D e^a \, , \\
F(J)^{ab}&= d \omega^{ab} + \frac{1}{2} [\omega,\omega]^{ab} \, ,
\end{align}
and $D X^a = d X^a + {\omega^{a}}_b X^b$ is the covariant exterior derivative with respect to rotations. We have the gauge transformations $\mathcal{A} \to \mathcal{A} + \mathcal{D} \alpha$ with $\mathcal{D}= d +[\mathcal{A}, \ ]$. This gives, decomposing $\alpha = \theta \Pi + \xi^a P_a + \lambda^a C_a + \Omega^{ab} J_{ab}$
\begin{equation}
\begin{aligned}
\delta_\alpha \mathcal{A} &= \delta_\alpha  n \ \Pi + \delta_\alpha e^a \ P_a + \delta_\alpha f^a \ C_a + \delta_\alpha \omega^{ab} \ J_{ab} \,   \\
& = \left(d \theta  + \lambda^a e_a - \xi^a f_a \right) \Pi + \left(D \xi^a + {\Omega^a}_b e^b \right) P_a \, \\
& + \left( D \lambda^a + {\Omega^a}_b f^b \right)C_a + D \Omega^{ab} J_{ab} \, . \label{gaugetrafo}
\end{aligned}
\end{equation}
It is convenient to introduce the one form $\Sigma= \lambda^a e_a - \xi^a f_a$ so that $n \to n + \Sigma$ under boosts.\newline
 One may derive classical Ward identities by considering the variation of the action
\begin{equation}
\delta S = \int \sqrt{g} \left( \pi^\mu  \delta n_\mu + t^\mu_a \delta e_\mu^a + b^\mu_a \delta f_\mu^a + \frac{1}{2} {S^\mu}_{ab} \delta {\omega_\mu}^{ab}\right) \, .
\end{equation}
from which one gets upon gauge variation
\begin{align}
& \frac{1}{\sqrt{g}} \partial_\mu \sqrt{g} \pi^\mu = 0 \, , \\
& \frac{1}{\sqrt{g}} D_\mu \sqrt{g} t^\mu_a + f_{\mu a} \pi^\mu = 0 \, , \\
& \frac{1}{\sqrt{g}} D_\mu \sqrt{g} b^\mu_a - e_{\mu a} \pi^\mu = 0 \, , \\
& \frac{1}{2 \sqrt{g}} D_\mu \sqrt{g} S^\mu_{ab} - t_{[ab]} - b_{[ab]} =0 , \\
& t^a_a + z \pi^\mu n_\mu + (z-1) b^a_a = 0 \, . 
\end{align}
In particular, if the boost current $b^\mu_a$ vanishes (that is the effective action is independent of $f_\mu^a$), then the boost Ward identity reads
\begin{equation}
\pi^a=0 \, ,
\end{equation}
We will come back to this identity in the last Section of the paper with regards to our original Lifschitz system.\\

\noindent It is important to notice that these Ward identities are in general different from the ones obtained by demanding diffeomorphism invariance. Making them compatible amounts to imposing a certain set of conditions on the gauge fields which we now discuss. These subtle points, once addressed, also allow to make contact with the geometric formulation of \cite{hartong2015gauging}. \newline

\noindent In gauging spacetime symmetries it is often the case that one chooses to impose a maximal set of curvature constraints on $\mathcal{F}$ in order to realize the spacetime algebra on a reduced set of fields. These constraints are also key in connecting the first order formalism with diffeomorphism invariance, since they allow to treat diffeomorphisms as being generated by a ($\mathcal{A}$-dependent) gauge transformation. Let us take as an example the Poincaré algebra generated by $P^a$ and $J^{ab}$. In this case one imposes the torsionless constraints $F(P)^a=0$. This has two consequences
\begin{itemize}
	\item The spin connection $\omega^{ab}$ can be expressed in terms of $e^a$ once the algebraic inverses $E^\mu_a$ are introduced.
	\item A diffeomorphism generated by the vector field $\xi^\mu$ may be interpreted as a $P$ transformation generated by the parameter $\xi^a = i_\xi e^a$. This is a consequence of the identity 
	\begin{equation}
	\mathcal{L}_\xi \mathcal{A}= i_\xi \mathcal{F} + \delta_{\alpha_\xi} \mathcal{A} \, , \ \ \alpha_\xi = i_\xi \mathcal{A} \, . \label{gaugetodiffeo}
	\end{equation}
	The l.h.s. of the equations is a diffeorphism acting on the connection, the r.h.s. a combination of a gauge transformation plus a spurious term $i+\xi \mathcal{F}$. The curvature constraints in this case make such a term drop out and one may re-express diffeomorphisms in terms of ($\mathcal{A}$-dependent) gauge transformations. In general one however has to deal with this term and show that, once the appropriate constraints are imposed, it can also be expressed as a gauge transformation.
\end{itemize}
The second point in extremely important in identifying the conserved currents of the theory. Being able to realize diffeomorphisms as gauge transformations assures that we may express the underlying stress tensor as a (linear combination of) the global currents.\newline
In Carrollian theories can also impose a similar torsion-less constraint 
\begin{equation}
F(\Pi)=F(P)^a=0 \, .
\end{equation}This allows to re-express the spin connection $\omega^{ab}_\mu$ and boost connection $f^a_\mu$ in terms of the fields $n_\mu$ and $e_\mu^a$. The splitting of the diffeomorphisms $\xi^\mu = \theta v^\mu + \xi^a E_a^\mu$ is also recovered once one imposes $i_v e^a = i_{E^a} n =0$. This is the approach used in \cite{bergshoeff2017carroll} to construct a version of Carrollian gravity. After solving the curvature constraints one finds the usual vielbein expression for $\omega^{ab}$, while
\begin{equation}
f_\mu^a = n_\mu v^\nu E^{\rho a} \partial_{[\nu} n_{\rho]}  + E^{\nu a} \partial_{[\mu} n_{\nu]} + S^{ab} e_{\mu b} \, ,
\end{equation} 
with $S^{ab}$ a symmetric tensor.\newline
In our case, however, we would like to define a geometry with nonvanishing $F(\Pi)$ in order to model external torsion. There are two sets of constraints that one may impose that achieve this.\\

\noindent The first is to set 
\begin{equation}
F(P)^a = F(C)^a = 0 \, . \label{Carr:constraint2}
\end{equation}
The first is the familiar vielbein constraint, which we use to determine the spin connection. We interpret the second as defining a Stueckelberg field $M$ for boosts, given by solving the equation
\begin{equation}
d M = f^a \wedge e_a \, .
\end{equation}
Existence of solution is guaranteed by \eqref{Carr:constraint2}, since in this case
\begin{equation}
d \left(f^a \wedge e_a\right)= D f^a \wedge e_a - f^a \wedge D e_a = F(C)^a \wedge e_a - f^a \wedge F(P)_a =0 \, .
\end{equation} 
This allows to define a one-form that transforms only under anisotropic translations $\hat{n}= n - M$. From which one finds 
\begin{equation}
F(\Pi)= d \hat{n} \, ,
\end{equation}
as an invariant curvature. To be precise, invariance follows from evaluating $\delta_\alpha \mathcal{F}=[\mathcal{F},\alpha]$, so that $F(\Pi)$ is gauge invariant only once the constraint \eqref{Carr:constraint2} is imposed. As a last remark, in the known examples of Carrollian theories, the gauge field $f^a$ drops out of the classical action, this means that we might as well set it to zero, rendering $M=0$ a viable solution. This is a particularly simple background that we will often use to extract physical predictions.\\

\noindent A similar situation, which we however do not analyze in depth in this article, is to go to a geometry without curvature but with nonvanishing torsion. This is given by the constraints
\begin{equation}
F(J)^{ab}=F(C)^{a}=0 \, .
\end{equation}
This also allows to define a Stueckelberg field for boosts by the integrability condition $f^a = D M^a$ for a zero form $M^a$\footnote{Notice that this solves the curvature constraints only if $F(J)^{ab}=0$ too, since $D D M^a \sim F(J)^{a b} M_b$.}. As before we can construct an invariant curvature
\begin{equation}
\bar{F} = d (n - M^a e_a) = F(\Pi) - M^a F(P)_a \, .
\end{equation}
The main difference is that now the field $n - M^a e_a$ transforms as follows
\begin{equation}
\delta_\alpha (n-M^a e_a)= d \theta - d (M^a \xi_a) \, .
\end{equation}
While these two sets of constraints are useful from the point of view of the descent equations, since the allow to define the invariant polynomical $P_n = F^n$ or $\bar{P}_n = \bar{F}^n$, they need to be analyzed in more detail to see how diffeomorphisms can be implemented through them. Since we will be using only the first set of constraints, we only give a detailed explanation regarding them.\newline

\noindent We thus analyze the constraints $F(P)^a=F(C)^a=0$. It proves useful to use the fact that we have a well defined vector field $v^\mu$ to decompose the possible generators of diffeomorphisms $\xi^\mu$. These fall into two categories
\begin{itemize}
	\item $\xi^\mu = \theta v^\mu$ , which one may call anisotropic diffeomorphisms. These will be adsorbed in the anisotropic translations of $n$.
	\item The complementary set. These may be expanded trough the inverse vielbein $E^\mu_a$ which is however not boost invariant. The defining property is then that they can be brought in the form $\xi^\mu = \zeta^a E_a^\mu$ by a boost.  
\end{itemize}
In the first case it is a simple matter to show that, if we further assume $i_v e^a=0$, that the diffeomorphism may be realized on $n$ and $e$ by a gauge transformation with $i_\xi \mathcal{A}= \alpha$ plus a further boost $\beta^a$ which fulfills $\theta i_v d n = \beta^a e_a$. This is always possible given the condition $i_v e^a=0$. All of these equations follow from inspecting \eqref{gaugetrafo}, \eqref{gaugetodiffeo}.

Vector fields which are not parallel to $v^\mu$ are more complicated to deal with. Even though one may implement the transformation in $e^a$ by the usual rule of identifying $\xi^a=\zeta^a$, since the repsective curvature vanishes, we also have conditions on $n$. The condition is extracted as before and we must assure that $i_\xi d n = \tilde{\beta^a} e_a$ to re-adsorb it as a boost. This means simply that $i_v \mathcal{L}_xi n =0$, which can be re-expressed as a condition on $v$ by recalling that $i_v n=1$ and thus $\mathcal{L}_\xi i_v n =0$. This leads to
\begin{equation}
i_v \mathcal{L}_\xi n = - i_{\mathcal{L}_\xi v} n \, .
\end{equation}
One solution to this equation is to restrict the group of allowed diffeomorphisms to the so called Carrollian diffeomorphisms \cite{Duval:2014lpa}. In our notation this means that $\mathcal{L}_\xi v=0$ this then assures that $\mathcal{L}_\xi n= \tilde{\beta}^a e_a$ which can be removed by a boost.\newline

\noindent The fact that we have an invariant curvature then is not surprising. In fact the restriction to Carrollian diffeomorphisms requires the introduction of a further connection $b_\mu$ which gives rise to an effective torsion at the geometric level \cite{Ciambelli:2018ojf}, we review such a fact in Appendix \ref{App:Carrdiff}.
Thus one concludes that the constraints imposed on the Carrollian curvatures allow to implement a special subset of \emph{Carrollian} diffeomorphisms as gauge transformations of the Carrollian connection.\newline

\noindent We conclude by pointing out that this first order formulation was given a spacetime interpretation in \cite{hartong2015gauging}. The idea is to use the vielbein postulates
\begin{equation}
\mathcal{D}_\mu n_\nu = \mathcal{D}_\mu e_\nu^a = 0 \, ,
\end{equation} 
together with their algebraic inverses, to fix the Christoffel symbols. Computing the commutator between covariant derivatives gives the following identifications between torsion and Riemannian curvature and the Lie algebra valued curvature $\mathcal{F}$:
\begin{equation}
\Gamma_{[\mu\nu]}^\rho = v^\rho F(\Pi)_{\mu\nu} + E^\rho_a F(P)^a_{\mu\nu} \, ,
\end{equation}
and
\begin{equation}
{R_{\mu\nu}}^{\alpha}_{\beta}= v^\alpha e_\beta^a F(C)^a_{\mu\nu} + E^\alpha_a e_\beta^b {F(J)_{\mu\nu}}^{a}_b \, .
\end{equation} 
One thus sees that imposing the flatness constraints amounts to having only torsion proportional to $v^\rho$ and Riemann curvature coming from the spin connection. We will henceforth use this in our formulas to identify $F(\Pi)_{\mu\nu}$ with the torsion $T_{\mu\nu}$.\newline
The one-form $M$ is introduced in this case as a tangent space field $M^a$ which can be used to make the Christoffel connection boost invariant, the two are related by $M= e_a M^a$.. For the complete expressions see \cite{hartong2015gauging}.\newline

\noindent Having introduced Carrollian geometry we move to the classification of possible anomalies in two and four dimensions.
\subsection{The consistency condition}
Let us start by recalling what the Wess-Zumino consistency condition is. Let $W[\mathcal{A}]$ be the effective action in the background of the gauge field $\mathcal{A}$. The presence of an anomaly amounts to a nontrivial variation of the effective action under gauge transformations
\begin{equation}
\delta_\alpha W[\mathcal{A}] = \textbf{A}_\alpha \, , \label{Cons:Anomaly}
\end{equation}
while $\textbf{A}_\alpha$ cannot be written as the gauge variation of a local functional of the external gauge fields. Since the gauge transformations form an algebra one should have that $[\delta_\alpha,\delta_\beta]= \delta_{[\alpha,\beta]}$ on functionals of the external fields. Applying this to \eqref{Cons:Anomaly} gives the celebrated consistency condition
\begin{equation}
\delta_\alpha \textbf{A}_\beta - \delta_\beta \textbf{A}_\alpha = \textbf{A}_{[\alpha,\beta]} \, .
\end{equation}
The solution to this problem is in general very hard, however a simple class of solutions can be found from the so-called descent equations, which allow to construct the anomaly polynomial starting from an invariant polynomial $P_{d+2}$ in two extra dimensions. The general Chern-Simons form is given by the familiar integral representation
\begin{equation}
CS_{d+1}[\mathcal{A}]= \frac{1}{d+1}\int_0^1 dt P_{d+2}[\mathcal{A}, \mathcal{F}_t, ... , \mathcal{F}_t] \, ,
\end{equation}
with $\mathcal{F}_t = d \mathcal{A}_t + \frac{1}{2}[\mathcal{A}_t,\mathcal{A}_t]$ and $\mathcal{A}_t = t \mathcal{A}$. The anomaly is obtained by the descent equation
\begin{equation}
\delta_\alpha CS[\mathcal{A}]= d \textbf{A}_\alpha \, . \label{WZ:chernsimons}
\end{equation}
While it may be tempting to directly write down a Chern-Simons form for the unconstrained Carroll algebra, but this turns out to be problematic in $d \neq 3$. The main problem is to define a nondegenerate bilinear form on the Lie algebra. The trouble here, as well as in other non-relativistic contexts, comes from finding some element $X$ which has a nonvanishing trace with $\Pi$, i.e. $tr \left(\Pi X\right) \neq 0$. 
One may do this by also gauging the dilatation generator. Then the Carroll algebra reduces to a truncation of an affine Kac-Moody algebra, for which Chern-Simons terms are known, see e.g. \cite{bonora1991affine}. In three spacetime dimensions the situation is ameliorated by the fact that the $SO(1,1)$ generator $J$ is a scalar or by considering further central extensions. This is however strongly dimension-dependent and tends to be at odds with the inflow mechanism. For a discussion of nonrelativistic Chern-Simons terms and related issues see e.g. \cite{papageorgiou2009chern,Hartong:2016yrf}.\newline
 The way we propose out of this conundrum is that, once we have imposed the curvature constraints $F(C)^a=F(P)^a=0$, one is able to explicitly write down an invariant polynomial by using the Stueckelberg gauge field. This allows us to recover, for example, the results of \cite{hofman2015warped,jensen2017locality} in two dimensions and to extend them to four.\newline
We will proceed as follows: we start by writing down the structure of the anomalies coming from the invariant polynomial $F(\Pi)^n$, then use the set of gauge fields to write down the most general Bardeen counterterms. These allow to show the mixed nature of the anomalies and to connect to previous results.\\

\noindent As a final remark, it is straightforward to see that the consistency conditions for the scaling symmetry (which amount to ask the anomaly to be dimensionless) cannot be satisfied by any local combination of fields for $z \neq 0$. This follows from the dimension assignments
\begin{align}
[n]&= z \, , \ \ [f]= -1 + z \, , \ \ [e]= 1 \, , \\
[\theta] &= z \, , \ \ [\lambda]= -1+z \, , \ \ [\xi]= 1 \, .
\end{align}
Thus in the following discussion we will restrict to the warped case $z=0$ only. It can be seen using the equations above that our final expression all have zero weight under the warped Lifschitz scaling and thus satisfy the consistency condition coming from scale invariance.

\noindent This restriction makes perfect sense from the perspective of our original computation in Section \ref{Sec: LIfschitz}: in such cases the relevant contributions to the Jacobian are regulator dependent and should be discarded. This is admittedly somewhat unpleasant from the perspective of the Carrollian algebra, since this has consistent realizations for all values of $z$. We have however not way around this obstruction as of yet.

\paragraph*{Two dimensions} We start by discussing warped theories in two spacetime dimensions. These have been object of quite a lot of interest through the last few years, see e.g. \cite{detournay2012warped,hofman2011chiral,hofman2015warped,castro2015warped,jensen2017locality}. In this case it is known that (in flat spacetime) the Carrollian group gets enhanced to a direct product $\rm Vir \times U(1)_{2 q^2}$ of a Virasoro times a $U(1)$ Kac-Moody algebra of the same chirality. The level $k=2q^2$ is fixed in terms of the anisotropic momentum $q$ of the fields. In this case one may naively expect a $U(1)$ anomaly from the Kac-Moody algebra, together with the diffeomorphism anomaly for the chiral Virasoro symmetry.\newline
We start from the invariant polynomial 
\begin{equation}
C_{4}(F) = \kappa F(\Pi)^2 \, ,
\end{equation}
and write down the Chern-Simons action
\begin{equation}
CS_3[n,M]= \kappa \int (n-M) \wedge F(\Pi) \, ,
\end{equation}
which leads to the anomaly
\begin{equation}
\textbf{A}_\alpha = \kappa \int \theta F(\Pi) \, .
\end{equation}
Notice that with this choice, the anomaly is purely in the anisotropic translations and takes the familar form of a chiral anomaly, which makes sense, since the underlying symmetry algebra is a Kac-Moody algebra. This is however not the whole story, since we may also introduce local (Bardeen) counterterms 
\begin{equation}
B_{M}= \tilde{\kappa} \int n \wedge M \, ,
\end{equation}
making use of the Stueckelberg field. Its gauge variation is
\begin{equation}
\delta_\alpha B_M = \tilde{\kappa} \int \left(\theta d M + \Sigma (n-M) \right)  \, ,
\end{equation}
this shifts the anomaly to the boost sector, in particular we may choose $\tilde{\kappa}=-\kappa$ to get
\begin{equation}
\textbf{A}_\alpha = \kappa \int \theta dn - \Sigma (n-M) \, .
\end{equation}
Notice that, even if we now impose $M=0$ and $dn =0$ we still have a boost anomaly coming from the volume element
\begin{equation}
\kappa \int  \lambda \ d Vol(\mathcal{M}) = \kappa \int  \lambda \ e \wedge n \, ,
\end{equation}
which was the resulting boost anomaly in \cite{jensen2017locality}.\\
To construct a Chern-Simons action we may also look for a three-form which satisfies \eqref{WZ:chernsimons}. The construction is carried out explicitly in Appendix  \ref{App:cons2d}, with the resulting term
\begin{equation}
CS[n,M,\rho]=\kappa \int n (dn -2 d M) \, .
\end{equation}
There are of course various other possibilities depending on the choice of counterterms, we show in Appendix \ref{App:cons2d} how to explicitly compute some solutions and comment on different choices of curvature constraints.
\paragraph*{Four dimensions} We can apply the exact same procedure to the four dimensional theory. In this case one may start from the invariant polynomial
\begin{equation}
P_6 = \kappa F(\Pi)^3 + \kappa_g F(\Pi) F(J)^{ab} F(J)_{ab} \, ,
\end{equation}
where the second term stand for a (possible) mixed translation-Lorentz anomaly. This gives a Chern-Simons term
\begin{equation}
\tilde{CS}[n,M,\omega]= \kappa \int (n-M) F(\Pi)^2 + \kappa_g \int (n-M)F(J)^{ab} F(J)_{ab} \, ,
\end{equation}
which gives the anomaly
\begin{equation}
\textbf{A}_\alpha = \kappa \int \theta F(\Pi)^2 + \kappa_g \int \theta F(J)^{ab} F(J)_{ab} \, . \label{Geo:Anomaly4d}
\end{equation}
Notice how this formally coincides with a chiral anomaly in four dimensions. In this case too we have a choice of Bardeen counterterm which allows to shift the anomaly into the boost and Lorentz sector given by
\begin{equation}
B_{M,\omega}= \tilde{\kappa}_1 \int n M \rho + \tilde{\kappa}_2 \int n M dn + \tilde{\kappa}_g \int (n-M) tr \left( \omega d \omega + \frac{2}{3} \omega^3 \right) \, .
\end{equation}
In particular $\tilde{\kappa}_1=-2 \kappa$, $\tilde{\kappa}_2=\kappa$ gives the anomaly we compute explicitly in Appendix \ref{App:cons2d}. While we could show explicit in the Lifschitz system that the first in the anomaly equation \eqref{Geo:Anomaly4d} in nonzero for a warped Lifschitz system, with contribution $\sim q^3$, we could not get the second term. It would be interesting to show under which circumstances this may arise, since it is bound to give interesting physical consequences similar to the chiral vortical effect. 
\paragraph{Further remarks} We conclude this part with some general remarks on our results.\newline
First, one has to keep in mind that these are actually the \emph{consistent} anomalies of the theory. The coefficients of the covariant anomalies can be found by defining expliticly the covariant current operators and work out exactly as in the case of abelian anomalies, see e.g. \cite{Landsteiner:2016led} for a recent review.\newline
Second, as usual, the coefficient $\kappa$ has to be determined from the microscopic theory by an explicit computation. We will see in what follows, however, that is should be proportional to $q^2$ in 2d and $q^3$ in 4d, since for free models one can always explicitly have anisotropic translations and boosts act as an internal symmetry under which the degrees of freedom transform projectively.\newline
Third, we have shown that a consistent anomaly may be defined for Carrollian theories, it is however known that none exists for Galileian ones. This is to be expected, since even dimensional Galileian theories are nothing but odd dimensional quantum field theories compactified over a null direction, and those have no continuous anomalies. On the other hand, Carroll theories may be defined simply by embedding the system into a null surfacce of the ambient spaceitime \cite{Afshar:2015wjm,Donnay:2019jiz,Ciambelli:2018ojf}. From this perspective the anomaly can be understood through the Callan-Harvey inflow mechanism \cite{Callan:1984sa}.
\subsection{Transport and warped anomalies}
Now that we have constructed the possible consistent anomalies and understood the functional dependence of their coefficients on the microscopic details of the theory, we ask ourselves what macroscopic features can they display. We will be interested in one point functions of the anisotropic momentum current $\pi^\mu$ in nontrivial gauge backgrounds. These are expected to be universal, as one may use the consistent anomaly to solve for their dependence on the state variables (for example chemical potentials and temperature). General methods to extract this kind of predictions from the knowledge of 't Hooft anomalies have seen a long story of development during the last years, see e.g. \cite{Son:2009tf,Jensen:2012jy,Jensen:2012kj,Jensen:2013kka,Jensen:2013rga,Jensen:2013vta,stone2018mixed,Loganayagam:2011mu,Loganayagam:2012pz,Loganayagam:2012zg}. In this section we will refrain to give a general analysis of the possible phenomena in warped theories, but rather limit ourselves to explain known results in two dimensions and give new predictions in two dimensions in some special cases. We will be interested in zero temperature effects only, although we will allow a nontrivial chemical potential $\Upsilon_\mu$ to be given to the anisotropic momentum current $\pi^\mu$. This enters the action through a coupling 
\begin{equation}
\int \sqrt{g} \left( \Upsilon_\mu \ \pi^\mu \right)\, , \label{Trs:coupling}
\end{equation}
taking $\Upsilon_\mu= \mathfrak{v} e_\mu^a t_a$ with $t_a$ the time direction, this corresponds to having the system at finite \emph{momentum density} in the anisotropic direction. This may be interpreted in an hydrodynamic framework by saying \cite{copetti2019anomalous} that the system is in a state with a finite velocity $\mathfrak{v}$ in the anisotropic direction. For the most part, we will leave $\Upsilon_\mu$ a generic constant and give a physical interpretation of its consequences after deriving the relevant results.\newline
Since in anomalous theories one is free to define various current operator, there is more than one way in general to characterize the physical response of the system. We will look at the following
\begin{description}
\item [Consistent currents] that are defined as functional derivative of the effective action $W[n,e,f,M]$ with respect to the external fields. They are not gauge invariant objects but their properties are simple to derive from the consistent anomaly of the theory. These where studied in part in \cite{copetti2019anomalous}.
\item [Covariant currents] that are defined by being gauge invariant even in the presence of external fields. Because of this their correlators are guaranteed to be gauge-independent. They can be constructed by explicitly computing the gauge variations of the consistent currents and subtracting it off through the appropriate Bardeen polynomials.
\end{description}
The main tool we are going to employ follows from the fact that we may see the coupling \eqref{Trs:coupling} as gauge equivalent to a configuration $n'$ of $n$ with
\begin{equation}
n'= n +d \theta \, , \ \ d \theta = a \, .
\end{equation}
The effective actions in the background of $n'$ and $n$ are then related by the anomaly equation\footnote{Such finite difference equation only holds for abelian transformations, which can be trivially integrated from their infinitesimal counterpart. In general what one gets on the r.h.s. is a generalization of the Liouville action, which arises from ``integrating" the anomaly polynomial to finite gauge transformations
\begin{equation}
W[\mathcal{A}']-W[\mathcal{A}]= \mathcal{W}_{\rm Anom} [g,\mathcal{A}] \, .
\end{equation}}
\begin{equation}
W[n',e,f,M]-W[n,e,f,M]= \textbf{A}_{\theta} \, ,
\end{equation}
since $n'= n + a$ the first term on the l.h.s. in the effective action in a background $n$ field with nontrivial ``chemical potential" $a$, which can be used to compute whatever correlator we may need, up to terms independent on $a$. This fixes the consistent currents up to $a$-independent terms. Examining the covariant currents fixes also these last terms since they must compensate the Bardeen polynomial.\newline
In two dimensions we find, after partial integration (we always use Bardeen counterterms to set the boost anomaly to zero)
\begin{equation}
W[n',e,f,M]-W[n,e,f,M]= -\kappa \int \Upsilon \wedge (n-M)
\end{equation}
which gives the consistent current $\pi^\mu$ to be 
\begin{equation}
\pi^\mu = \kappa \epsilon^{\mu\nu} a_\nu = \kappa \left(v^\mu E^\nu \Upsilon_\nu - E^\mu v^\nu \Upsilon_\nu \right) \, ,
\end{equation}
decomposing $a_\mu= n_\mu \Upsilon_v + e_\mu \gamma$ we thus find that the momentum density $\pi= e_\mu \pi^\mu$ gets a nontrivial expectation value form the boost anomaly
\begin{equation}
\pi = \pi(n,e,f)- \kappa \Upsilon_v \, ,
\end{equation}
as pointed out by Jensen \cite{jensen2017locality} while a nontrivial spectral flow changes the value of $\pi_v$ (which is the holomorphic Kac-Moody current in the language of \cite{hofman2015warped}) to
\begin{equation}
\pi_v = \pi_v(n,e,f) + \kappa \gamma \, ,
\end{equation}
which matches previous expressions after taking $\kappa= k/4\pi = q^2/2\pi$ and taking into account factors of $1/2\pi$ from the CFT normalization of operators. One may also construct the covariant current $\pi_{cov}^\mu$ by adding the Bardeen Polynomial
\begin{equation}
B^\mu= \kappa \epsilon^{\mu\nu}(n-M)_\nu \, ,
\end{equation}
to $\pi^\mu$. This allows to fix $\pi(n,e,f)$ in a background with $M_\mu=0$ by equating it to $-B^\mu e_\mu$ thus
\begin{equation}
\pi(n,e,f)=\kappa \, , \ \ \pi_v(n,e,f)=0 \, ,
\end{equation} 
furthermore the transport coefficients above get multiplied by a factor of two for the covariant current.\newline
In four dimensions the situation is similar, only that now we find
\begin{equation}
W[n',e,f,M]-W[n,e,f,M]= -\kappa \int \Upsilon \wedge (n-M)F(\Pi) \, ,
\end{equation}
which now implies a nontrivial response of $\pi^\mu$ to external torsion $F(\Pi)$. We have
\begin{equation}
\pi^a = 2 \kappa \epsilon^{abc} \left(\Upsilon_v  T_{bc} + \Upsilon_b G_c \right) + \pi^a(n,e,f) \, , \label{Transport:Streda}
\end{equation}
while
\begin{equation}
\pi_v = 2 \kappa \epsilon^{abc} \Upsilon_c T_{ab} + \pi_v(n,e,f) \, .
\end{equation}
Now the Bardeen polynomial is
\begin{equation}
B^\mu = 2 \kappa \epsilon^{\mu\nu\rho\sigma} (n-M)_\nu F(\Pi)_{\rho\sigma} \, ,
\end{equation}
if $M=0$ this gives, equating the gauge dependent part of the consistent current to minus the Bardeen polynomial
\begin{equation}
\pi^a(n,e,f)= 2 \kappa \epsilon^{abc} T_{bc} \, ,
\end{equation}
which is precisely the result of \cite{copetti2019anomalous} once we identify the coefficient of the consistent anomaly as
\begin{equation}
\kappa = \frac{s q^3}{96 \pi^2} \, ,
\end{equation}
which is the right anomaly for a chiral fermion. The covariant responses, on the other hand, acquire a factor of $3$ with respect to the consistent ones.\newline
Let us also speculate a bit on the effect of the mixed gravitational anomaly in four dimensions. In the chiral case one gets a temperature dependent response in a thermal state to an external gravito-magnetic field, which is dubbed the chiral vortical effect \cite{Landsteiner:2011cp}. In this case the situation should be completely analogous, with however the gravito-magnetic field coming from the spatial metrix $h_{\mu\nu}$ only. The proof of this statement is however more involved, since one needs to use methods akin to those of \cite{stone2018mixed} to put the system on a nontrivial background geometry. In particular, this should be a thermal warped geometry which should be studied in detail on its own. \newline
Finally, let us give a somple example of the consequences of \eqref{Transport:Streda} on the macroscopic physics. We consider the case in which $\Upsilon_t = \mathfrak{v} \neq 0$, that is, we are in a system at finite velocity in the anisotropic direction. Suppose also that the torsion $F(\Pi)$ is located in the 2 dimensional plane perpendicular to $v^\mu$. The nonvanishing Burgers vector $b^\mu= \epsilon^{\mu\nu\rho\sigma}t_\nu F(\Pi)_{\rho\sigma}$ is then also in the $v^\mu$ direction.
What we find by applying \eqref{Transport:Streda} is that, whenever the Burgers vector and the velocity overlap, anisotropic momentum density $\pi$ is formed. This momentum density is localized on dislocations and is created by the Carrollian ``fluid" passing through them, since it is proportional to the velocity $\mathfrak{v}$.

\section{Warped CFTs}
In this last section we present some significant examples of free warped CFTs to show explicitly that the $z \to 0$ limit of our fermionic theory falls into such a class. It is helpful to recall the (classical) Ward identities for a Warped Carrollian system, which can be copied from Section \ref{Sec:Carroll} by setting $z=0$:
\begin{align}
& \frac{1}{\sqrt{g}} \partial_\mu \sqrt{g} \pi^\mu = 0 \, , \\
& \frac{1}{\sqrt{g}} D_\mu \sqrt{g} t^\mu_a + f_{\mu a} \pi^\mu = 0 \, , \\
& \frac{1}{\sqrt{g}} D_\mu \sqrt{g} b^\mu_a - e_{\mu a} \pi^\mu = 0 \, , \\
& \frac{1}{2 \sqrt{g}} D_\mu \sqrt{g} S^\mu_{ab} - t_{[ab]} - b_{[ab]} =0 , \\
& t^a_a - b^a_a = 0 \, . 
\end{align}
In all of the examples which are studied, furthermore, it can be checked that $f_\mu^a$ drops out of the Lagrangian, so that the current $b^\mu_a$ vanishes. Then we have:
\begin{align}
& \frac{1}{\sqrt{g}} \partial_\mu \sqrt{g} \pi^\mu = 0 \, , \\
& \frac{1}{\sqrt{g}} D_\mu \sqrt{g} t^\mu_a = 0 \, , \\
& e_{\mu a} \pi^\mu = 0 \, , \\
& \frac{1}{2 \sqrt{g}} D_\mu \sqrt{g} S^\mu_{ab} - t_{[ab]} =0 , \\
& t^a_a = 0 \, . 
\end{align}
In particular we will concentrate on the anisotropic translation current $\pi^\mu$ and its physics.
\subsection{Free examples}
Let us move on to discuss free warped fermionic theories following \cite{hofman2015warped,jensen2017locality}. Their construction is two dimensional, however we will present a realization of the Carrollian Clifford algebra which allows a slight generalization to higher dimensions. This procedure has the advantage that warped Lagrangians may be immediately written by using the Dirac operator $\slashed{D}$ for the Carroll algebra.\newline

\noindent The main obstacle that we need to overcome to generalize such definitions is to define a Clifford algebra for Carrollian theories. To the author's knowledge the definition of Clifford algebras for nonrelativistic theories is not a completely understood topic, but some examples can be found essentially by trial and error. A guiding principle may be to consider the embedding of the (euclidean) non-relativistic group into $SO(d+1,1)$, which has a natural Clifford algebra. The non-relativistic algebra is then obtained by Inonu-Wigner contraction.\newline

\noindent A more simplistic perspective, yet sufficient for the case at hand, is to fix the basic anticommutation relations by requiring them to close on a (twice-contravariant) invariant tensor. For the Carroll group this is just $v^\mu v^\nu$. This is not, however, the complete construction, since consistency of the whole Clifford algebra (in particulat having Hermitean conjugates defined, since we will have nihilpotent generators) requires adding one further generator, which in our case will be denote by a charge conjugation matrix $\Gamma^{\mathcal{C}}$.

The relevant Clifford algebra turns out to be spanned by matrices $\Gamma^a$, $\Gamma^n$ satisfying
\begin{equation}
(\Gamma^n)^2= \mathbb{I} \,  , \ \ \{\Gamma^a, \Gamma^b\}= 0 \, , \ \ \{\Gamma^n,\Gamma^a\}=0 \, ,
\end{equation}
together with a charge conjugation matrix $\Gamma^\mathcal{C}$ which satisfies\footnote{This is needed since nihilpotent matrices cannot be Hermitian.}
\begin{equation}
\mathcal{C} \Gamma^a \mathcal{C}^{-1}= (\Gamma^a)^\dagger\equiv \tilde{\Gamma}^a \, , \ \ \ \{\Gamma^\mathcal{C},\Gamma^n \} =0 \, , \ \ \ \left(\Gamma^{\mathcal{C}}\right)^2= -\mathbb{I} .
\end{equation}
Notice that this is different from the Galileain Clifford algebra of Levy-Leblond \cite{levy1967nonrelativistic}, for which one should take
\begin{equation}
(\Gamma^\pm)^2= 0 \, , \ \ \{\Gamma^+,\Gamma^-\}=2 \mathbb{I} \,  , \ \ \{\Gamma^a, \Gamma^b\}= 2 h^{ab} \mathbb{I} \, , \ \ \{\Gamma^n,\Gamma^a\}=0 \, ,
\end{equation}
which comes from the null embedding of the Galileian group. The reason for this difference is essentially found in the difference in invariant tensors, $v^\mu v^\nu$ for the Carroll case and $h^{\mu\nu}$ for the Galileian.\newline

\noindent Since the $\Gamma^a$ are nihilpotent, we may split them in two by two block form. Let us assume that the blocks are of the same size, so that spinors of the two may transform in the same representation of $SO(d-1)$. A representation of the Carrollian Clifford algebra is then given by
\begin{equation}
\Gamma^n = \begin{pmatrix}
\mathbb{I} & 0 \\
0 & -\mathbb{I}
\end{pmatrix} \, , \  \ \Gamma^a = \begin{pmatrix}
0 & \gamma^a \\
0 & 0
\end{pmatrix}
\, , \ \ \Gamma^\mathcal{C}= \begin{pmatrix}
0 & \mathcal{C} \\
-\mathcal{C} & 0
\end{pmatrix} \, ,
\end{equation}
with $\gamma^a$ Clifford matrices of $SO(d-1)$ and $\mathcal{C}$ the charge conjugation matrix of such reduced Clifford algebra.
In this way we can define the boost and rotation generators as follows
\begin{align}
C^a &= \frac{1}{4}[\Gamma^a,\Gamma^n]= -\frac{1}{2} \begin{pmatrix}
0 & \gamma^a \\
0 & 0
\end{pmatrix} = - \frac{1}{2} \Gamma^a \, , \\
\Omega^{ab} &= \frac{1}{4} [\Gamma^{[a}, \tilde{\Gamma}^{b]}] = \begin{pmatrix}
\gamma^{ab} & 0 \\
0 & - \gamma^{ab} 
\end{pmatrix} \, ,
\end{align}	
with $\gamma^{ab}=\frac{1}{4} [\gamma^a,\gamma^b]$.
Notice that, whithout charge conjugation, no rotation generators are present in the algebra. There are now two possibilities: one is to consider single component\footnote{By ``single compoenent" we mean fermion representations of the reduced rotation group, in our case either none or SO(1,2).} fermions $\varphi$ which do not transform under boosts, by using the projector $P_{-}= \frac{1}{2}\left(\mathbb{I}-\Gamma^n\right)$. The other is to consider two component fermions $\Psi=(\chi,\varphi)$ of which only $\chi$ transforms under boosts $\chi \to -\frac{1}{2}\lambda_a \gamma^a \varphi$. In this framework the Dirac operator $\bar{\Psi} i \slashed{D} \Psi$ with $\slashed{D}= \Gamma^a D_a + \Gamma^n D_v$ is by construction boost invariant as can be explicitly checked.
The first of these representations is commonly called the ``Weyl" representation, while the second case goes under the name of ``bc" representation.

\paragraph{Weyl representation}
The first representation only allows for a nontrivial kinetic term of the form $\bar{\varphi} D_v \varphi$ which is the component version of $\bar{\Psi} \slashed{D} P_{-} \Psi$, with $\bar{\Psi}= \Psi^\dagger \left(\Gamma^\mathcal{C}\right)^{-1}$. If $\varphi$ is complex we may also add a (dimensionless) mass term $q \bar{\varphi}\varphi$, so that
\begin{equation}
S_{Weyl}= \int d^d x \sqrt{g} \left(i \bar{\varphi} D_v \varphi + q \bar{\varphi}\varphi \right) \, ,
\end{equation}
being $g_{\mu\nu}= e_\mu^a e_\nu^b \eta_{ab} + n_\mu n_\nu$. This action is boost invariant since each term is. The warped scaling symmetry acts by $x_a \to \lambda x_a$ , $\varphi \to \lambda^{-(d-1)/2}\varphi$. Notice also the presence of the spurionic symmetry $x_\mu \to r x_\mu$, $\varphi \to r^{-(d-1)/2} \varphi$ , $q \to r^{-1} q$ as we expect. The dynamical equations read
\begin{equation}
i D_v \varphi + q \varphi = 0 \, ,
\end{equation}
which basically force $\varphi$ to be a plane wave with fixed momentum. It is interesting to also define the conserved currents, in particular the anisotropic momentum current $\pi^\mu$ which turns out to be
\begin{equation}
\pi^\mu = \frac{1}{2} i \bar{\varphi} v^\mu D_v \varphi \, ,
\end{equation}
which of course satisfies the boost Ward identity $\pi^a=0$ and, on-shell
\begin{equation}
\pi^\mu = q v^\mu \bar{\varphi} \varphi \, .
\end{equation}
This fixes how $q$ appears in correlators involving the current.

\paragraph{BC representation}
On the other hand, for the ``bc" system one can expand the Dirac operator as
\begin{equation}
i \bar{\Psi} \slashed{D} \Psi = i \bar{\varphi} \gamma^a D_a \varphi + i (\bar{\chi} D_v \varphi + \bar{\varphi} D_v \chi) \, ,
\end{equation}
and boost invariance can be explicitly checked. This theory also allows for a mass term $q \bar{\Psi} \Psi$. Furthermore, in contrast to the Weyl system a Majorana condition may be imposed, which fixes $\chi$ and $\varphi$ to be real. The action then reads
\begin{equation}
S_{bc}= \int \sqrt{g} \left( i \varphi^T \mathcal{C}^{-1} \gamma^a D_a \varphi + 2 i \chi^T \mathcal{C}^{-1} D_v \varphi  + 2 q \chi^T \mathcal{C}^{-1} \varphi \right) \, .
\end{equation}
The scaling symmetry now is implemented by $x_a \to \lambda x_a$ , $\varphi \to \lambda^{-(d-2)/2}\varphi$ , $\chi \to \lambda^{- d/2} \chi$. As before, this theory too has a spurionic symmetry involing $q$, under which $x_\mu \to r x_\mu$ , $\Psi \to r^{-(d-2)/2} \Psi$ , $q\to r^{-1} q$. The dynamical equations read
\begin{align}
i D_v \varphi + q \varphi &= 0 \, , \\
i \gamma^a D_a \varphi + \left( i D_v -q \right) \chi & = 0 \, .
\end{align}
The second equation may be thought as determining $\chi$ given the isotropic profile of $\varphi$. The anisotropic momentum current is now given by a more lengthy formula
\begin{equation}
\pi^\mu = v^\mu \left(i \left\{ \bar{\chi} D_v \varphi + \bar{\varphi} D_v \chi \right\}  - \frac{1}{2} \mathcal{L}\right)    + i E^\mu_a \left( \bar{\varphi} \gamma^a D_v \varphi \right)
\end{equation}
after substituting the solution to the dynamical equations $\varphi = e^{i q x_v} \tilde{b} \, , \ \ \chi = e^{-i q x_v} b - \frac{i}{2 q}  e^{i q x_v} \gamma^a D_a \tilde{b} $ one finds 
\begin{equation}
\pi^\mu = \frac{q}{2} v^\mu \tilde{b} b \, .
\end{equation}
Which is also proportional to $q$.
\paragraph{Projective transformations}
In both cases boosts and anisotropic translations may be reabsorbed as internal transformations. To see this, define for the Weyl fermion\footnote{The careful reader will realize that these are just the solutions to the dynamical equations. Here however we keep the modes to depend on $x_v$ too, so that it should be seen as a change of variables.}
\begin{equation}
\varphi= e^{i q x_v} \eta \, ,
\end{equation} 
and, for the ``bc" theory
\begin{align}
\varphi = e^{i q x_v} \tilde{b} \, , \ \ \chi = e^{-i q x_v} b - \frac{i}{2 q}  e^{i q x_v} \gamma^a D_a \tilde{b} \, ,
\end{align}
the variable $\eta$ now transforms projectively under boosts and anisotropic diffeomorphisms (since $\varphi$ was invariant), that is
\begin{equation}
\eta \to e^{- i q (\theta + \lambda_a x^a)} \eta \, ,
\end{equation}
in a representation of charge $q$.
In the same way, the fields $b$, $\tilde{b}$ transform as a chiral multiplet $Z=(b,\tilde{b})$ as
\begin{equation}
Z \to e^{i q \Gamma^n (\theta + \lambda_a x^a )} Z \, .
\end{equation}
the actions then become
\begin{equation}
S_{Weyl}= i\int \sqrt{g} \bar{\eta} D_v \eta \, ,
\end{equation}
and
\begin{equation}
S_{bc}= i\int \sqrt{g} Z^T \mathcal{C}^{-1} D_v Z \, ,
\end{equation}
hence the name ``bc" theory. In both cases one may think of the symmetry as a chiral $U(1)$ transformation acting on this new set of variables. In this way, one can get an intuitive explanation on why we get precisely such a form for the anomaly. Following the standard lore about abelian anomalies, this has to be proportional to $q^2$ in two dimensions, since it can be computed by a diagram with two $\eta$ current insertions, and, according to the same reasoning, to $q^3$ in three dimensions. This can be checked using the on-shell expressions for the current operators. In two dimensions this has already been verified by using OPE techniques \cite{castro2015warped}, while we leave the detailed computation in higher dimensionality for the future.\newline

\noindent Let us also note, following \cite{jensen2017locality}, that these theories (both Weyl and bc) admit an infinite number of exactly marginal deformations obtained by sandwiching arbitrary powers of $i D_v$ in boost invariant bilinears (e.g. $\mathcal{O}_n = i \bar{\Psi} \left(i D_v\right)^n \slashed{D} \Psi$ ). Their presence makes the theory borderline non-local. However it is interesting to point out that they all break the spurionic symmetry and thus may not emerge in approaching the warped limit from a Lifschitz theory, which only has one marginal parameter. 

\subsection{Emergence of Carrollian symmetry}
It is also natural to ask ourselves whether a further symmetry emerges in the warped limit which allows for nontrivial anomalies. The answer is affirmative and can be (classically) shown as follows. The anisotropic momentum current has  components $\pi^a=\pi^\mu e_\mu^a= e_\mu^a \frac{1}{\sqrt{g}}\frac{\delta S}{\delta n_\mu}$ given by \cite{copetti2019anomalous}
\begin{equation}
\pi^a \sim \varphi^T \mathcal{C}^{-1} \gamma^a \nabla_v \varphi \, ,
\end{equation}
we may define $\beta^a =  \mathcal{C}^{-1} \gamma^a$ which are symmetric imaginary matrices. The equations of motion for our system read\footnote{Here we work in flat spacetime for simplicity.}
\begin{equation}
i \gamma^a \nabla_a \varphi = s q \left( i \nabla_v /q\right)^{1/z}\varphi \, ,
\end{equation}
excluding zero modes of both operators these can be equivalently written as
\begin{equation}
i \nabla_v \varphi = q \left(i \gamma^a \nabla_a/ s q \right)^{z} \varphi \, .
\end{equation}
Taking the warped limit this gives, modulo the subtleties above
\begin{equation}
i \nabla_v \varphi = q \varphi \, , \label{carrolllimit}
\end{equation}
which imply the equation
\begin{equation}
\pi^a \sim q \varphi^T \beta^a \varphi = 0 \, , 
\end{equation}
since $\beta^a$ are symmetric and $\varphi$ is a Grassmann variable. We thus have the classical Ward identity
\begin{equation}
\pi^a = 0 \, , \label{Lif:WIboost}
\end{equation}
Which ensures Carrollian boost symmetry to emerge in the warped limit. It is worthwhile to ask which of the two free theory representations is consistent with the warped limit of the Lifschitz system. A naive answer would be the ``Weyl'' one, since there is only one field involved. However this is not possible because the scaling dimensions do not match. The only anwer consistent with the warped Lifschitz scaling is given by the bc system. It would be interesting to understand whether the Weyl system can also be defined through an appropriate warped limit.

\noindent Let us conclude this section by giving an intuitive understanding of the parameter $q$ in the warped limit, since it appears to be important in defining the anomalies of the theory. It is known in the literature that Carrollian particles ``cannot move" \cite{bergshoeff2014dynamics}. This is because the Hamiltonian is a central element of the algebra and sets the energy to a fixed quantity (i.e. it cannot be raised by kinetic energy). In our approach the role of the energy is taken on by the anisotropic momentum, and the Carrollian particle is in a ``frozen" plane wave state with momentum $q$. This cannot be changed by scattering it with other Carrollian particles and thus acts and a fixed charge for the system, much akin to an internal Abelian symmetry.\\

\noindent Finally, by confronting equation \eqref{carrolllimit} with the dynamical equations for warped fermions, we see natural to identify the respective marginal couplings given by $q$ and the warped mass term.

\section{Conclusions and open questions}
In this paper we have analyzed the warped limit of fermionic Lifschitz theories from various perspectives. In particular we have shown that a nontrivial anomaly polynomial can arise in such limit and matched these predictions with the known free theory examples. Our conclusions predict the presence of universal transport properties regarding the anisotropic momentum current, for which we have given an effective theory interpretation if $z=0$. Furthermore, one should be able to generalize our arguments to explain the results of \cite{Landsteiner:2016stv}. From that perspective the nontrivial viscosity should be explained along the lines of the chiral vortical effect.\newline
There are however a number of open questions which should be addressed in the future
\begin{itemize}
	\item  We have computed the anomaly for Lifschitz fermions by the Fujikawa procedure, can this be extended to the warped case directly? If so, does the ``chirality'' matrix $\Lambda$ play a role similar to $\Gamma_{d+1}$ in the standard case? The Fujikawa computation is often most simply carried out by mapping the problem to SUSY-QM. What is the super-symmetric quantum mechanical model corresponding to the warped fermions? In a similar spirit, SUSY-QM requires a precise understanding of the fermionic representations of non-relativistic groups. We feel that this requires further investigation.
	\item In our computations we have disregarded interactions. These are known to be hard to incorporate for warped theories, based on dimensional analysis only. Can one use such interactions to flow from/to different values of $z$ and extend our results?
	\item Our solutions to the consistency conditions rely strongly on the presence of curvature constraints. Without those, no Chern-Simons term can be defined due to the absence of an invariant bilinear form. Such a form is available for Carroll groups only in three dimensions. Can one find a different embedding of the (warped) Carroll group which allows for a nondegeneate bilinear form? If so what is its physical interpretation?
\end{itemize}

\paragraph*{Acknowledgements}
This work is supported by FPA2015-65480-P and by the Spanish Research Agency (Agencia Estatal de Investigación) through the grant IFT Centro de Excelencia Severo Ochoa SEV-2016-0597. The work of C.C. is funded by Fundaci\'on La Caixa under ``La Caixa-Severo Ochoa'' international predoctoral grant. the author would like to thank Karl Landsteiner and Eric Bergshoeff for discussions and comments on the draft. He also would like to thank the organizers of the ``Effective Theories of Quantum Phases of Matter'' workshop at NORDITA, where part of this work was presented.

\appendix
\section{Expansion of the regulated Jacobian}\label{app:GravJac}
In this first Appendix we give the expansion of the regulator and the computation of the gravitational contributions to the anomaly in four dimensions. Recall that we have defined
\begin{equation}
\mathcal{R}= A^\dagger A \, , \ \ \ A=\frac{i \gamma^a \nabla_a/q}{\Lambda_1} + s \frac{\left(i \nabla_v/q \right)^{1/z}}{\Lambda_2}\, ,
\end{equation}
in this appendix we will drop factors of $q$ for ease of notation, they are straightforward to re-enstablish. The regulator then can be computed by expanding
\begin{equation}
\frac{1}{\Lambda_1^2}\gamma^a \gamma^b \nabla_a \nabla_b + \frac{i s}{\Lambda_1 \Lambda_2}  [\gamma^a \nabla_a,\left(i \nabla_v \right)^{1/z}] +\frac{1}{\Lambda_2^2} \left(i \nabla_v \right)^{2/z} \, ,
\end{equation}
the last term does not give any interesting space-time dependence and is the source for the anisotropic Gaussian term. We then need to compute various commutators, these are given by the formula
\begin{equation}
[\nabla_\mu,\nabla_\nu]= -T^\rho_{\mu\nu}\nabla_\rho + R_{\mu\nu}^{ab} J_{ab}\, ,
\end{equation}
with $J_{ab}$ the relevant rotation generator.
\begin{equation}
\begin{aligned}
&\gamma^a \gamma^b \nabla_a \nabla_b= \nabla_\perp^2 +  \frac{i}{2}\epsilon^{abc} \gamma_c [\nabla_a,\nabla_b]= \\
&= \nabla_\perp^2 - i \frac{i}{2}  \epsilon^{abc} \gamma_c T_{ab} \nabla_v + \frac{1}{4} \epsilon^{abc}\epsilon_{efg} \gamma_c \gamma^g {R_{ab}}^{ef} \, ,
\end{aligned}
\end{equation}
the first term will contribute the the Gaussian integral, the second to the torsional anomaly while the third may be further massaged into 
\begin{equation}
\frac{1}{4} \epsilon^{abc}\epsilon_{efg} \gamma_c \gamma^g {R_{ab}}^{ef}= \frac{1}{2} R - \frac{i}{2} \epsilon^{abc} R_{abcf} \gamma^f \, .
\end{equation}
We also have the commutator
\begin{equation}
\gamma^a[\nabla_a , \nabla_v]= -\gamma^a G_a \nabla_v + \frac{i}{2}\epsilon_{efg} \gamma^a \gamma^g {R_{a v }}^{ef} \, .
\end{equation}
which appears in $i s  [\gamma^a \nabla_a,\left(i \nabla_v \right)^{1/z}]$ repeatedly. The leading term in the plane wave expansion is given by $1/z$ times such commutator, multiplied by the plane wave momentum $ k_v^{1/z-1}$.\newline
Passing to a plane wave basis for the computation of the trace and rescaling the momenta as $k_a=\Lambda_1 u_a$, $k_v= \Lambda_2^z v$ one is led to the following integral expansion
\begin{equation}
\begin{aligned}
J(\theta)&= \frac{\theta}{(2\pi)^d} \Lambda_1^{d-1} \Lambda_2^z \int d^{d-1}u_a \int d x_v \left(\nabla_v + i \Lambda_2^z v \right) \exp\left(-u_a u^a - v^{2/z}\right) \times \, \\
& \times \exp\left(i \frac{u^a \nabla_a}{\Lambda_1} - \frac{\nabla_\perp^2- R}{\Lambda_1^2} +\frac{i}{\Lambda_1^2}\epsilon^{abc} \gamma_a T_{bc} (\nabla_v + i v \Lambda_2^z )  +        \right. \\
 &\left. +\frac{s}{\Lambda_1 \Lambda_2}\sum_k^{1/z} c_k \nabla_v^k \gamma^a\left\{ R_{av}^{ef}\gamma_{ef} +  G_a (\nabla_v +i v \Lambda_2^z)   \right\} \left(\nabla_v + i v \Lambda_2^z\right)^{1/z-k}    \right. \\
 &\left. \frac{\left\{ (\nabla_v + i v \Lambda_2^z)^{2/z} - \Lambda_2^2 v^{2/z} \right\}}{\Lambda_2^2} \right) \, .
\end{aligned}
\end{equation}
Notice that all terms with indeces $a,b,c...$ in the non-Gaussian part decay at least as $1/\Lambda_1$ or $1/\Lambda_2$ so that our expansion will terminate at the third order in four dimensions an at the second order in two dimensions. The torsional part is particularly simple, since the $T_{ab}$ contribution already comes in as $\Lambda_1^{-2}$. In this case the only contribution comes together with the highest weight term in $G_a$ to give the integral in the main text. Recall also that, in order to preserve the spurionic symmetry, $\Lambda_2= q^{1/z-1} \tilde{\Lambda}_2$.  \newline
In the absence of torsion, one could hope to find further contributions from the Riemann tensor in four dimensions, however there seems to be no such contribution from our system. In any case the evaluation of gravitational anomalies using the Fujikawa technique is known to be cumbersome and this question should be further studied by Supersymmetric methods.

\section{Warped transport from Kubo formulas in the Lifschitz theory}\label{App:Kubo}
Here we briefly compute the response to anisotropic velocity (i.e. chemical potential) in the Kubo formalism. We work at zero temperature and with non-vanishing chemical potential $a_b$, so that the propagator reads\footnote{$a_b$ couples to the current $\pi^a$ by definition.}
\begin{equation}
S_F = \left(\gamma^a k_a + s (k_v/q)^{1/z} + a_b \gamma^b k_v\right)^{-1}\,
\end{equation}
we can to compute the two point functions of two $\pi$ currents at nonvanishing anisotropic momentum $q_v$ with the propagator above. This is given by the Feynman graph
\begin{equation}
\langle \pi^a(q_v) \pi^b(-q_v) \rangle = -\frac{1}{2} \int \frac{d^3 k_a d k_v}{(2\pi)^4} tr \left[(k_v +q) \gamma^a S_F(q_v +k) k_v \gamma^b S_F(k) \right] \, ,
\end{equation}
we evaluate the odd part of the trace to first order in $a_b$ and $q_v$ finding
\begin{equation}
= \epsilon^{abc} a_c i q_v \frac{2 s}{\pi^3} \int_0^\infty d k k^2 \int_0^\infty d k_v k_{v}^{1/z +2} \times \frac{1}{\epsilon(k)^4} \, ,
\end{equation}
being $\epsilon(k)$ the dispersion relation. Changing variables to $k_v=q p^z$ and then to polar coordinates $(p,k)= x(\sin(\phi),\cos(\phi)$ we find
\begin{equation}
=  q^3 \epsilon^{abc} a_c i q_v \frac{2 s}{\pi^3} z \int_0^{\pi/2}\cos(\phi)^2 \sin(\phi)^{3z} \times \int_0^\infty x^{3z-1} \, ,
\end{equation}	
the angular integral gives $\frac{1}{2} B(3/2, (3z+1)/2)$ being $B(a,b)$ the Euler beta function while we need to regulate the radial integral
\begin{equation}
z \int_0^\infty x^{3z-1}= z \lim_{\alpha \to 0} x^{3z-1} \exp(-\alpha x)= \lim_{\alpha \to 0} z \Gamma(3z) \alpha^{-3z} \, ,
\end{equation}
taking the limit $z \to 0$ before the limit $\alpha \to 0$ gives the finite result we are looking for, as in the computation of the torsional anomaly, it is a UV divergent term to give a finite result in the warped limit. Putting everything together and including a factor of $1/2$ for the Majorana fermions we get
\begin{equation}
\lim_{q_v \to 0} \frac{\langle \pi^a(q_v)\pi^b(-q_v)\rangle}{i q_v} = \frac{s q^3}{24 \pi^2} \epsilon^{abc} a_c + O(a^2)\, .
\end{equation}
\section{Carroll manifolds and Carrollian diffeomorphisms}\label{App:Carrdiff}
In this appendix we briefly review the definition of a Carrollian manifold. We mostly follow \cite{Ciambelli:2018ojf,ciambelli2018flat}. One defines a Carrollian manifold $\mathcal{C}= \mathbb{R} \times \mathcal{S}$ by taking local coordinates $(t,\textbf{x})$ and specifying the set of allowed Carrollian diffemorphisms
\begin{equation}
t'=f(t,x) \, , \ \ \textbf{x}'=\textbf{g}(\textbf{x}) \, ,
\end{equation}
according to this subset of diffeomorphisms ordinary exterior derivatives on $\mathcal{C}$ do not transform as forms as the $\textbf{x}$ and $t$ derivatives mix since
\begin{equation}
\partial_i'= J^i_j \left( \partial_i + \partial_i t \partial_t \right)\, , \ \ J^i_j = \frac{\partial x^i}{\partial x'^j} \, .
\end{equation}
This can be solved by introducing a connection $b_\mu$ so that 
\begin{equation}
\partial_{i} \to \partial_{i} + b_i \partial_t \equiv D_i \, ,
\end{equation}
which adsorbs the anomalous transformation by
\begin{equation}
b'_i = b_i - \partial_i t \, ,
\end{equation}
the consequence of this is that we have introduced an effective torsion into our system as
\begin{equation}
[D_i,D_j]= (db)_{ij} \partial_t \, .
\end{equation}
To make contact with the Lie algebra formulation $d t = d \theta$ and $b_i = (n-M)_i$. Thus a torsional gauge field arises naturally when dealing with Carrollian manifolds. One then needs to embed this construction in a more general coordinate independent setting. The first step is to define Carrollian diffeomorphisms. These are generated y the vector fields
\begin{equation}
\xi = \alpha(t,\textbf{x}) \partial_t + \xi^i(\textbf{x}) \partial_i \, ,
\end{equation}
in this case too one may distinghish two families of vector fields. The first has only a $\partial_t$ component, which is a statement robust under Carrollian diffeomorphisms, while the second can be put in the form $\xi^i(\textbf{x})\partial_i $ in an appropriate Carrollian coordinate system. The fact that $\xi^i$ does not depend on $t$ reads
\begin{equation}
\mathcal{L}_{\partial_t} \xi^i =0 \, .
\end{equation}
Which is the condition we have imposed in the main text (of course, in general, $\partial_t$ is represented by the invariant vector field $v$). We thus see that the curvature conditions we have used in our approach are equivalent to working on a Carrollian manifold.
\section{Consistency conditions}\label{App:cons2d}
In this Appendix we explicitly show how to derive the solution to the consistency conditions once the curvature constraints $Df=De=0$ are imposed. Throughout we introduce the two form $\rho= f^a \wedge e_a$ to streamline the notation.
\paragraph*{Two dimensions} We start in two dimensions with an Ansatz for the Chern-Simons term, we write
\begin{equation}
CS[n,M,\rho]= \kappa_1 \int n dn + \kappa_2 \int n \rho \, ,
\end{equation}
its variation is
\begin{equation}
\delta_{\theta,\Sigma} CS[n,M,\rho]= \kappa_1 \int d (\theta dn) - \kappa_1 \int d (\Sigma n) + (2\kappa_1-\kappa_2) \int n d\Sigma \, ,
\end{equation}
where we have used $\Sigma \rho = 0$ in two dimensions.  Taking $\kappa_2=-2\kappa_1$ thus gives a solution
\begin{equation}
CS[n,M,\rho]=\kappa \int n (dn -2 \rho) \, .
\end{equation}
This gives an anomaly
\begin{equation}
\textbf{A}_{\alpha}=\kappa \int \left[\theta (dn - 2 f\wedge e) - \xi n \wedge f + \lambda n \wedge e   \right] \, .
\end{equation}
We can also verify directly that this is a valid solution to the consistency conditions. Taking into account the scaling dimensions of the various fields we are left with the following terms in the warped limit
the following terms are possible
\begin{align}
A_\Pi= dn + b f \wedge e \, , \\
A_P= n \wedge f \, , \\
A_C= n \wedge e \, ,
\end{align}
with $b$ some real constant. We thus parametrize
\begin{equation}
\delta_\alpha W = \kappa_1 \int A_\Pi + \kappa_2 \int A_P + \kappa_3 \int A_C \, .
\end{equation}
First we take variations with respect to $\theta_1$ and $\xi_2$. The parameters do no vary themselves in this scheme and the variations commute. Then
\begin{equation}
\kappa_1 \int \theta_1 \delta_{\xi_2} \left( dn + b f \wedge e \right) - \kappa_2 \int \xi_2 \delta_{\theta_1}(n \wedge e ) = 0 \, ,
\end{equation}
the left term gives $- \theta_1 d(\xi_2 f) + \theta_1 f \wedge d \xi_2$ while the right term gives $\xi_2  d \theta_1 \wedge f$. The two must be equal up to a total derivative. Rewriting $- \theta_1 d (\xi_2 f) = -d(\theta_1 \xi_2 f) + d\theta_1 \xi_2 f$ and imposing the curvature constraint $df=0$ gives a total derivative $d(\theta_1 \xi_2 f)$ if 
\begin{equation}
\kappa_1(1+b) = \kappa_2 \, , \ \ b= 0 \, .
\end{equation}
We now do the same but with the variation with respect to $\theta_1$ and $\lambda_2$. The calculation is the same as before, with the replacement $f \to e$ and a minus sign in the first term since there is no minus in the transformation. The solution then gives 
\begin{equation}
\kappa_1(1+b)= -\kappa_3 \, .
\end{equation}
using $\kappa_1$ as a variable we have
\begin{equation}
\delta W = \kappa_1 \int \left(\theta (dn +b f\wedge e) + (1+b)\zeta n \wedge f - (1+b)\lambda n \wedge e \right) \, .
\end{equation}
To fix $b$ we must impose the final consistency condition with translation $\xi_1$ and boost $\lambda_2$. This should create a nonvanishing $\theta_{12}=- \xi_1 \lambda_2$. Let us compute
\begin{equation}
\begin{aligned}
& \kappa_1(1+b) \int \lambda_2 \delta_{\zeta_1}(n\wedge e) +\kappa_1 (1+b)  \int \zeta_1 \delta_{\lambda_2} (n \wedge f)= \\
= & \kappa_1 (1+b) \int \left( \lambda_2 \left[- \zeta_1 f \wedge e + n \wedge d \zeta_1 \right] - \zeta_1 \left[\lambda_2 e \wedge f + n \wedge d \lambda_2    \right]  \right)= \\
= & -\kappa_1 (1+b) \int -\zeta_1 \lambda_2 \left( dn - 2 f \wedge e \right) \,
\end{aligned}
\end{equation}
having integrated by parts $ n (\xi_1 d \lambda_2 + \lambda_2 d \xi_1)= \xi_1 \lambda_2 dn - d(n \xi_2 \lambda_1)$. Thus our last equation reads 
\begin{equation}
\kappa_1(dn + b f \wedge h)= -\kappa_1(1+b)(dn -2 f\wedge e) \, ,
\end{equation}
which has only the solution $b=-2$. Calling then $\kappa_1=\kappa$ we have our final result
\begin{equation}
\textbf{A}_\alpha= \kappa \int \left[\theta (dn - 2 f\wedge e) - \xi n \wedge f + \lambda n \wedge e   \right] \, .
\end{equation}
\paragraph*{Four dimensions}
We now repeat the same discussion in the four dimensional case. We start from an ansatz for the Chern-Simons for (we follow the notation introduced in the body of the paper)
\begin{equation}
CS[n,M\rho]= \kappa \left( a_1 n dn^2 + a_2 n dn \rho + a_3 n \rho^2 + a_4 M \rho^2 \right) \, ,
\end{equation}
we will require
\begin{equation}
\delta_\alpha CS[n,M,\rho]= d \textbf{A}_\alpha \, .
\end{equation}
The mixed gravitational term, which is proportional to $\kappa_g$ can be seen to fullfill these conditions immediately. The $\theta$ variation is seen to give a total derivative
\begin{equation}
\delta_\theta CS[n,M,\rho]= d \theta \left( a_1 dn^2 + a_2 dn \rho + a_3 \rho^2 \right) \, ,
\end{equation}
the coefficients $a_i$ are fixed by requiring that boosts fullfill the same condition. The computation is actually made quite easier by the introduction of the parameter $\Sigma$ to account for it. We find
\begin{equation}
\begin{aligned}
\delta_\Sigma CS[n,M,\rho] &= d\left( (2a_1+a_2) \Sigma n dn + (2a_3+a_2) \Sigma n \rho + 2a_4 \Sigma M \rho \right) \, \\
&+ (3a_1+a_2) \Sigma dn^2 +2(a_2+a_3)\Sigma dn \rho +(a_3 + 3 a_4) \Sigma \rho^4 \, ,
\end{aligned}
\end{equation}
thus we find $a_2=-3a_1$, $a_3= 3 a_1$ , $a_4=-a_1$ and we absorb $a_1$ into $\kappa$. Finally we have the anomaly
\begin{equation}
\begin{aligned}
\textbf{A}_\alpha &= \kappa \int \theta\left( dn dn - 3 dn \rho + 3 \rho^2 + \right) + \kappa_g \int \theta F(J)^{ab} F(J)_{ab} \, \\
& - \kappa \int \Sigma \left( n dn - 3 n \rho + 2 M \rho \right) \, .
\end{aligned}
\end{equation}
There are two Bardeen counterterms for boosts
\begin{equation}
B_{M,\omega}= b_1 \int n M \rho + b_2 \int n M dn= b_1 B^{(1)}_M + b_2 B^{(2)}_M \, ,
\end{equation}
plus the usual one for Lorentz symmetry. We have
\begin{align}
\delta_\alpha B^{(1)}_M &= \int - \theta \rho^2 + \Sigma \left( -2 n \rho + M \rho + dn M \right) \, , \\
\delta_\alpha B^{(2)}_M &= \int - \theta dn \rho + \Sigma\left(-n dn + 2 M dn - n \rho\right) \, ,
\end{align}
asking the boost anomaly to be cancelled is solved by $b_1=2 \kappa$, $b_2=-\kappa$ which can be checked by substituting the lines above in the computation of the anomaly. The $\theta$ part of $\textbf{A}_\alpha$ then becomes
\begin{equation}
\kappa \int \theta \left(dn^2 -(3+1)dn \rho +(3-2)\rho^2 \right)= \kappa \int \theta R(\Pi)^2 \, .
\end{equation}
\paragraph*{Different choices of constraints}
We can also construct invariant polynomials by choosing the ``zero curvature" constraints
\begin{equation}
F(J)^{ab}=F(C)^a=0 \, ,
\end{equation}
these give rise to a geometry in which only torsion is present. In this case the invariant polynomial is given by
\begin{equation}
\bar{P}_n = \kappa \bar{F}^n \, ,
\end{equation}
with $\bar{F}= F(\Pi) - M_a F(P)^a$. We focus on two dimensions  for simplicity, the results extend to four dimensions. The computation of the anomaly gives
\begin{equation}
\textbf{A}_\alpha= \kappa \int \left(\theta - M^a \xi_a \right) \bar{F} \, , 
\end{equation}
we have, as before, one Bardeen counterterm at our disposal
\begin{equation}
B_M= \tilde{\kappa} \int n \wedge M \, ,
\end{equation}
whose variation gives 
\begin{equation}
\delta_\alpha B_M = -\tilde{\kappa} \int \Sigma (n-M) + \theta d M - M^a \xi_a dn \, ,
\end{equation}
which can be used to make the anomaly equivalent to the Chern-Simons term computed in the first part of the Appendix, modulo terms which vanish with $M$. In particular, we can recover the same boost anomaly by choosing $dn=M=0$.
\bibliography{AnomTrans}{}
\bibliographystyle{JHEP}

\end{document}